# Pedestrian wayfinding behavior in a multi-story building: a comprehensive modeling study featuring route choice, wayfinding performance, and observation behavior

Yan Feng*

Department of Transport & Planning, Civil Engineering and Geosciences Faculty, Delft University of Technology, y.feng@tudelft.nl

Dorine C. Duives

Department of Transport & Planning, Civil Engineering and Geosciences Faculty, Delft University of Technology, d.c.duives@tudelft.nl

**Abstract**

This paper proposes a comprehensive approach for modeling pedestrian wayfinding behavior in complex buildings. This study employs two types of discrete choice models (i.e., MNL and PSL) featuring pedestrian route choice behavior, and three multivariate linear regression (MLR) models featuring the overall wayfinding performance and observation behavior (e.g., hesitation behavior and head rotation). Behavioral and questionnaire data featuring pedestrian wayfinding behavior and personal information were collected using a Virtual Reality experiment. Four wayfinding tasks were designed to determine how personal, infrastructure, and route characteristics affect indoor pedestrian wayfinding behavior on three levels, including route choice, wayfinding performance, and observation behavior. We find that pedestrian route choice behavior is primarily influenced by route characteristics, whereas wayfinding performance is also influenced by personal characteristics. Observation behavior is mainly influenced by task complexity, personal characteristics, and local properties of the routes that offer route information. To the best of our knowledge, this work represents the first attempt to investigate the impact of the same comprehensive set of variables on various metrics feature wayfinding behavior simultaneously.

## 1. INTRODUCTION

Understanding pedestrian wayfinding behavior in buildings under normal and evacuation conditions is vital to pedestrian safety. Pedestrian wayfinding in a building refers to the decision-making process related to (1) choosing a route, (2) acquiring information from the environment, and (3) accordingly navigating to a destination (Shields and Boyce, 2000). These three levels of behavior are important elements of effective and efficient wayfinding (Ruddle and Lessels, 2006). Finding one's way inside of buildings can be difficult, especially in multi-story buildings. Studies showed that the wayfinding difficulty increases in multi-story buildings due to the increasing complexity of the network layout and movements across multiple vertical levels (Feng et al., 2022a; Kuliga et al., 2019).

A number of studies have investigated wayfinding behavior from either experimental or modeling perspectives (e.g., Cao et al., 2018; Duives and Mahmassani, 2012; Feng et al., 2022a; Hölscher et al., 2007). From an experimental point of view, traditionally, behavioral data featuring pedestrian wayfinding behavior have been collected using field experiments (e.g., Cao et al., 2018; Dong et al., 2020; Fang et al., 2010; Hölscher et al., 2007; Kobes et al., 2010b; Kuliga et al., 2019) and surveys (e.g., Duives and Mahmassani, 2012; Galea et al., 2017; Shiwakoti et al., 2017). Although these methods provide valuable information regarding pedestrian wayfinding behavior, these methods also have certain constraints, such as limited experimental control of external variables, privacy concerns, and ethical and financial restrictions (Feng et al., 2021a). Due to the high level of experimental control, the high granularity of the data, and the ability to immerse participants in realistic scenarios without exposing them to risk, Virtual Reality (VR) is more and more frequently adopted to collect pedestrian behavioral data (e.g., Feng et al., 2021a; Lin et al., 2019;

* Corresponding author

Lovreglio et al., 2021; Schrom-Feiertag et al., 2017; Silva et al., 2013; Suzer et al., 2018). Studies have demonstrated that VR is capable of collecting valid pedestrian wayfinding behavioral data (e.g., Ewart and Johnson, 2021; Feng et al., 2021b, 2022a; Kinateder et al., 2018; Lin et al., 2019).

From a modeling point of view, existing studies often simplified pedestrian wayfinding behavior by modeling only one of the above-mentioned metrics. Most pedestrian route choice models predominantly account for infrastructure and route characteristics for route choice through outdoor environments (e.g., Fossum and Ryeng, 2022; Sevtsuk and Basu, 2022). Wayfinding performance and observation behavior studies predominantly focused on the development, calibration, and validation of models that can simulate the movement dynamics of pedestrians (e.g., Blue and Adler, 2001; Dietrich and Köster, 2014; Helbing and Molnár, 1995; Hoogendoorn et al., 2014). Wayfinding performance and observation behavior are rarely translated into mathematical models. When studied, predominantly statistical analyses are presented featuring the impact of one particular factor on wayfinding performance, the number of hesitations or gazing behavior. Interestingly, although these three metrics are closely related, the three pedestrian behaviors are studied by three completely different communities. In addition, most works feature a limited subset of independent variables; often the impact of one factor on one type of behavior. Comprehensive models starring a multitude of variables are lacking in general. As a result, the insights on indoor route choice, wayfinding performance, and observation behavior are disjointed and incoherent. Moreover, few insights existed regarding the combinatory impact of the factors that are identified in the wayfinding behavior literature.

This paper aims to model pedestrian wayfinding behavior in a more comprehensive fashion. We hypothesize that the same set of personal, infrastructure, and route factors affect indoor pedestrian wayfinding behavior on three levels, namely route choice, wayfinding performance, and observation behavior. This paper employs two types of discrete choice models (i.e., MNL and PSL) featuring pedestrian route choice behavior, and three multivariate linear regression (MLR) models featuring the overall wayfinding performance and observation behavior (e.g., hesitation behavior and gazing behavior). Data from a VR experiment featuring four different wayfinding assignments with increasing complexity is adopted, including (1) a within-floor wayfinding assignment, (2) a between-floor wayfinding assignment (i.e., across the horizontal and vertical level), (3) a more complex between-floor wayfinding assignment, and (4) an evacuation assignment. Quantitative data featuring pedestrian route choice, wayfinding performance, and observation behavior was collected via the VR system. Qualitative data regarding personal characteristics and experience of the virtual environment are collected via a questionnaire. This unique VR data set allows us to model pedestrians' wayfinding behavior more comprehensively and identify the combinatory impact of the independent variables.

This paper adds to the literature on pedestrian wayfinding study by comprehensively modeling the wayfinding behavior of pedestrians in multi-story buildings for the first time. The route choice behavior is modeled by means of two commonly used discrete choice model approaches. At the operational level, the average wayfinding performance, the gazing behavior, and the hesitation behavior of the participants are modeled. In both cases, a wide array of infrastructure, route, and personal characteristics are included in the variable sets. For the first time, this approach allows us to describe pedestrian wayfinding behavior in complex buildings integrally. Moreover, it allows us to determine whether the same set of characteristics affects pedestrian route choice, wayfinding performance, and observation behavior, thus observing the combinatory impact and relative impact of the characteristics.

The structure of this paper is organized as follows. Section 2 reviews the state-of-the-art research on pedestrian wayfinding behavior. Accordingly, the VR experiment and data collection procedures of Feng et al. (2022a) are introduced in section 3. Section 4 presents the modeling methodology, where the model structure and model search strategy are identified. Section 5 presents the four distinct models, namely a route choice model, a wayfinding performance model, a hesitation model, and a gazing model. The discussion of the model findings is presented directly after the presentation of the modeling results in the same section. This paper ends with a summary of the main conclusions and suggestions for future research direction in section 6.

## 2. STATE-OF-THE-ART WAYFINDING STUDIES IN BUILDINGS

To understand pedestrian wayfinding behavior in buildings, three levels of metrics have been identified in the literature as critical components for effective and efficient wayfinding and evacuation (Ruddle and Lessels, 2006), including (1) decision making (e.g., route and exit choice), (2) wayfinding performance (e.g., wayfinding speed), and (3) observation (e.g., head rotation and hesitation). In this section, an overview of the literature on these three components of pedestrian wayfinding is provided. Before doing so, first the experimental methods used in previous endeavors to study pedestrian wayfinding behavior are discussed. Accordingly, we provide an overview of studies considering pedestrian route choice, wayfinding performance, and observation behavior.

### 2.1 Studying wayfinding behavior through experiments

To understand pedestrian wayfinding behavior, previous studies have employed different experimental methods to collect pedestrian behavioral data. Three types of experimental methods have often been employed, including field observations, surveys, and Virtual Reality experiments (Feng et al., 2021a).

Field observations often take place in existing buildings or temporary setups in a laboratory. Often cameras are employed to record pedestrian movements in natural and realistic settings. Several studies have investigated pedestrian exit choice behavior during wayfinding under emergency conditions using field observations (Cao et al., 2018; Fang et al., 2010; Fridolf et al., 2013; Heliövaara et al., 2012; Kobes et al., 2010a; Zhu and Shi, 2016). The advantages of field observation are that pedestrian movements are recorded in the most undisturbed settings. However, the location and scale of the observations are often limited and there are constraints in controlling external factors (Feng et al., 2021a). Due to the above-mentioned limitations, only a few studies adopted this method to study pedestrian wayfinding performance using location-track apps or cameras (Aksoy et al., 2020; Hölscher et al., 2007; Kuliga et al., 2019). More recently, with the development of eye-tracking techniques, more studies have begun to employ eye-tracking devices to investigate pedestrian gaze behavior during wayfinding (Afrooz et al., 2018; Bae et al., 2020; Dong et al., 2020; Ohm et al., 2017; Wang et al., 2019).

Surveys adopt pre-defined questions to investigate pedestrian choice behavior in past-event scenarios or hypothetical scenarios. The majority of studies have used surveys to investigate pedestrian route and exit choices during evacuations (Benthorn and Frantzich, 1999; Duives and Mahmassani, 2012; Galea et al., 2017; Haghani and Sarvi, 2016; Shiwakoti et al., 2017) and normal conditions (Veeraswamy et al., 2011). Surveys provide high experimental control to design predetermined questions. However, most survey studies employed the format of text and images, which inherently limit the realistic representation of complex environments and dynamic features of pedestrians' movement. Yet, the validity of survey results is still being questioned regularly (Feng et al., 2021a).

Compared to traditional experimental methods, VR has the advantage of higher experimental control and higher accuracy in collecting participants' movement data. Using VR experiments, a large number of studies investigated pedestrian route and exit choice behavior during wayfinding in both normal and emergency conditions in different types of buildings (Andree et al., 2015; Duarte et al., 2014; Feng et al., 2022b; Kobes et al., 2010a; Lin et al., 2019; Lin et al., 2020; Silva et al., 2013; Suzer et al., 2018; Vilar et al., 2014b). Meanwhile, due to the advantage of collecting data with a high level of granularity, wayfinding performance has been extensively studied using VR experiments, including total travel time (Kalantari et al., 2022; Lin et al., 2019; Nam et al., 2015; Schrom-Feiertag et al., 2017; Shi et al., 2021a; Suzer et al., 2018), total travel distance (Duarte et al., 2014; Lin et al., 2019; Meng and Zhang, 2014; Nam et al., 2015; Schrom-Feiertag et al., 2017), and average walking speed (Feng et al., 2022a; Lin et al., 2019; Zhang et al., 2021). Besides collecting movement trajectories, VR provides possibilities to collect other behavioral data, such as gazing data, which can be used to analyze pedestrian observation behavior during wayfinding. For instance, Zhang et al. (2021) and Feng et al. (2022a) investigated pedestrian head rotation change during wayfinding, which is the result of looking around in the environment. Duarte et al. (2014) and Feng et al. (2022a) studied pause behavior to understand the information acquisition process during wayfinding. Moreover, several studies analyzed the area of interest (AOI) during wayfinding in indoor settings using eye movement data (e.g., Cai et al., 2018; Feng et al., 2022a; Schrom-Feiertag et al., 2017; Tian et al., 2019).

### 2.2 Wayfinding behavior in buildings

#### 2.2.1 Research on pedestrian route choice

Pedestrian route choice behavior has been investigated through experiments and modeling. Most experimental studies focused on pedestrian exit choice behavior under emergencies in different types of buildings, including hotels (Kobes et al., 2010a; Vilar et al., 2014b), campus buildings (Fang et al., 2010; Feng et al., 2021b; Zhu and Shi, 2016), transport hubs (Galea et al., 2017; Haghani and Sarvi, 2016; Lin et al., 2020; Shiwakoti et al., 2017; Suzer et al., 2018), hospitals (Silva et al., 2013), museums (Lin et al., 2019), and company headquarters (Duarte et al., 2014). Most of the above-mentioned studies focused on pedestrian exit choice instead of route choice and the experimental scenarios were relatively simple (e.g., simplified environment, movement on a single level).

Concerning modeling studies, route choice models predict the probability of choosing one particular route from a discrete set of alternative routes, i.e., route set. Here, the underlying assumption is that the characteristics of a route, the person, the network, and the physical environment impact the likelihood that a pedestrian adopts a specific route. The literature on modeling pedestrian route choice behavior is summarized in Appendix 1, which identifies the type of population, the number of participants, the type of model, and the categories of characteristics included per study. This overview shows that most research features the route choice of adults in outdoor environments, predominantly urban areas. In most studies, convenience samples are used which are recruited through newspaper ads, invitations distributed in a certain region, or invitations distributed via email lists. The larger data samples (>5,000) mentioned in Appendix 1 all feature third-party datasets that were collected for other purposes. Often PSL and MNL models are used to determine the impact of route characteristics on pedestrian route choice. In particular, travel distance and the number of turns along the route are most often included in the model. The second most considered factor features infrastructure characteristics, which include design characteristics of a street at link level, ranging from the presence of benches and stairs to the type of pavement. In some studies, land use was considered, in particular the presence of shops, windows, and restaurants at street level.

Surprisingly, only a limited number of studies consider personal characteristics, such as age, gender, and education level, in the estimated route choice models. Moreover, very few indoor route choice models have been estimated. To the author's knowledge, 5 route choice models have been presented. Cheung and Lam (1998) adopted a binary logit model featuring the choice of stairs versus escalators in mass rapid transit stations. Tsukaguchi and Matsuda (2002) and Tsukaguchi and Ohashi (2007) studied route choice behavior around and in a large shopping mall in Japan. In the latter case, a binary logit model was also adopted to identify the probability of adopting a route. Stubenschrott et al. (2014) studied the route choice behavior in and around the U2 subway station in Zurich. Here, a simple utility maximation scheme was adopted. Lastly, Werberich et al. (2015) used a physics model to describe the route choice of pedestrians through a corridor network.

In sum, there are few insights concerning the type of routes that pedestrians prefer when walking indoors with multiple floors. Moreover, we have a limited understanding regarding whether and to what extent personal characteristics influence pedestrians' route choices.

### 2.2.2 Research on wayfinding performance

The literature on pedestrian wayfinding performance is summarized in Appendix 2. Research on wayfinding performance is quite rich and predominantly adopts field observations and VR experiments in which participants are asked to walk from A to B. Real-life public spaces, such as parks (e.g., Mackay Yarnal and Coulson, 1982; Malinowski and Gillespie, 2001), museums (e.g., Lin et al., 2019), campus buildings (e.g., Feng et al., 2022a), malls (e.g., Zhang and Park, 2021), hospitals (e.g., Kuliga et al., 2019; Pouyan et al., 2021), or virtual environments that resemble these real-life spaces are often adopted as a topic of study. Besides that, mazes are a popular topic of study (e.g., Chen et al., 2023; Shi et al., 2021b). Most studies research wayfinding movement along a horizontal plane. Since 2007, in a few studies, vertical movements are introduced as well, amongst which by Hölscher et al. (2007), Kuliga et al. (2019), and Feng et al. (2022a). These studies show that vertical movements involving staircases tend to disorient participants.

From the onset, the focus of most wayfinding performance studies has been to identify the impact of infrastructure factors to enhance performance. The color, style, and level of detail of visual cues in the infrastructure design, route guidance systems, maps, and augmented reality-like systems were the main subjects of study (e.g., Goldiez et al., 2007; O'Neill, 1992; Shi et al., 2021b; Soh and Smith-Jackson, 2004; Vilar et al., 2014a). The existing research found that the impact of those cues and their ability to optimize wayfinding performance, seems to be dependent on the complexity of the specific task at hand, the characteristics of the participant, their wayfinding strategy, and the network layout.

In recent years, a growing body of literature has investigated the impact of personal characteristics on pedestrian wayfinding performance. Especially gender is often studied, where male participants are regularly found to outperform female participants, i.e., males are faster and more efficient (e.g., Chen et al., 2009; Malinowski and Gillespie, 2001). However, some conflicting findings have also been reported. For instance, the study of Suzer et al. (2018) and Wang et al. (2019) showed that there is no difference in wayfinding performance between male and female participants. Regarding age, Lee and Kline (2011) and Wang et al. (2019) found that younger adults perform better in wayfinding tasks than older adults. In some studies, participants' previous experience in the building and cognitive abilities (e.g., path planning and spatial awareness) were explicitly accounted for (e.g., Kuliga et al., 2019; Morganti et al., 2007). In general, participants with better spatial awareness can find their way better, as do participants that are more familiar with a building (Li et al., 2019).

When analyzing how these studies present their results, we find that most studies are limited to trend analyses and statistical tests on the mean and variance of test samples. The literature on wayfinding especially describes the impact of a wide array of factors on wayfinding performance separately in great detail. Less attention is given to the interaction between different factors. To the authors' knowledge, only Li et al. (2019), Kuliga et al. (2019), and Kato and Takeuchi (2003) studied the conjoint impacts of multiple independent variables on wayfinding performance.

#### 2.2.3 Research on observation behavior

Compared to the studies of route choice and wayfinding performance, pedestrian observation behavior is rarely studied. The major challenge here is to collect sufficient data that can feature observation behavior. Only a few studies have investigated pedestrian observation behavior (e.g., hesitation, gazing) during wayfinding using new tracking technologies. For instance, Zhang et al. (2021) investigated the impact of the turning angle of architectural layout on pedestrian head rotation during wayfinding, while Feng et al. (2022a) analyzed the difference in head rotation among wayfinding tasks with varying complexity. Regarding hesitation behavior, Duarte et al (2014) compared hesitation behavior in evacuation scenarios with either static or dynamic signs, and Suzer et al. (2018) investigated the impact of environmental lighting conditions on hesitation behavior in an airport. Regarding gazing behavior, Cai et al. (2018) and Tian et al. (2019) investigated the gaze distribution and eye fixation under the condition with different types of emergency signs, while Wang et al. (2019) investigated the impact of age and gender on eye fixation. The literature on pedestrian observation behavior is summarized in Appendix 3. To the authors' knowledge, none of these works have attempted to model observation behavior.

## 3. VIRTUAL REALITY EXPERIMENT

Pedestrian wayfinding trajectory data in a multi-story building was collected using a VR study conducted from 27$^{th}$ November to 18$^{th}$ December 2019. Ethical approval was obtained from the Human Research Ethics Committee of the Delft University of Technology (Reference ID: 944). This section presents a detailed description of the VR experiment.

### 3.1 Experiment design

A VR research tool - WayR was created to collect pedestrian wayfinding data under normal and emergency conditions in a complex building. WayR features a virtual building that consists of three intermediate floors and one exit floor with eight emergency exits. Each floor comprises two main corridors, five staircases, and five elevators. Figure 1 shows the abstract layout of the virtual building. Meanwhile, WayR supports free navigation and automatically records pedestrian movement trajectories. Our previous work established the validity and usability of using WayR to study pedestrian wayfinding behavior. The detailed development and evaluation of WayR can be found in Feng et al. (2022a).

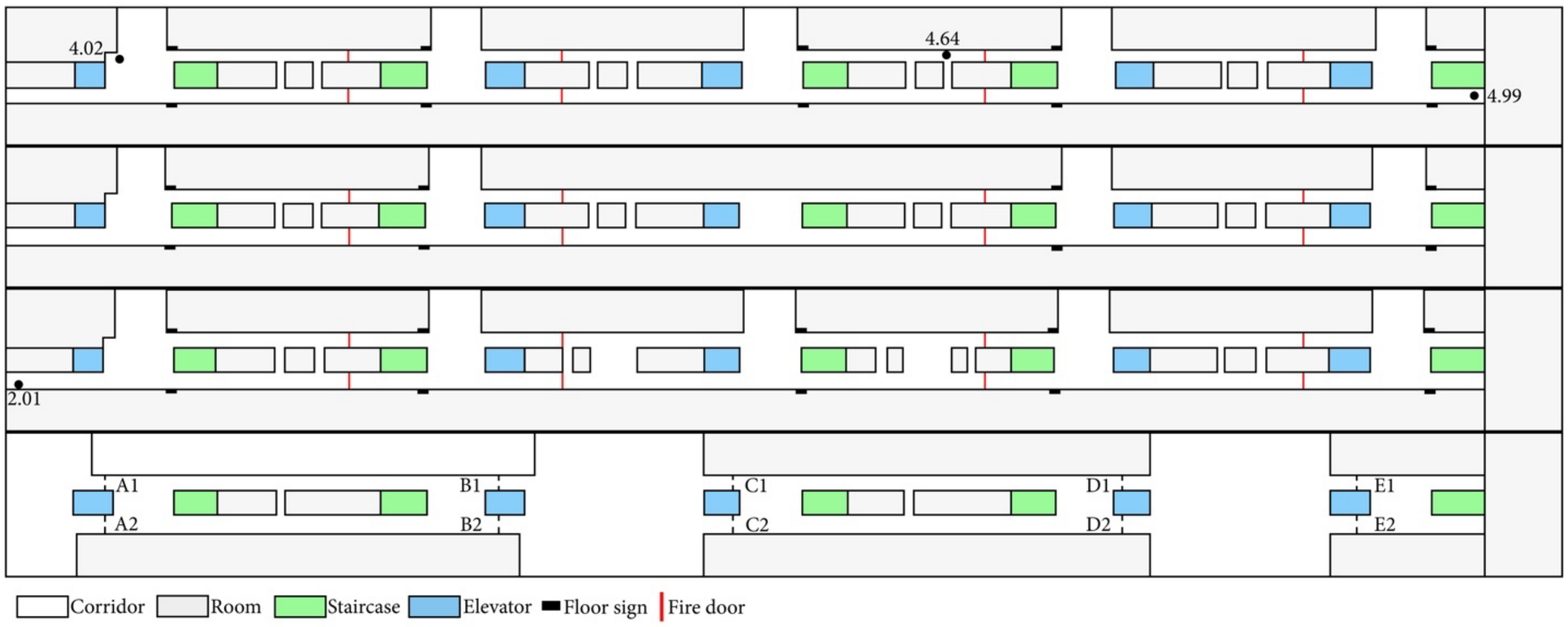

Figure 1. The abstract layout of the multi-level building (A1-E2 are exits).

The current experiment aims to investigate pedestrian wayfinding behavior in a complex building with various complexities, namely (1) a within-floor wayfinding assignment, (2) a between-floor wayfinding assignment (i.e., across the horizontal and vertical level), (3) a more complex between-floor wayfinding assignment, and (4) an evacuation assignment. Accordingly, four wayfinding assignments with increasing complexity were deliberately designed by applying WayR. Firstly, participants are asked to perform three wayfinding tasks under normal conditions, namely find their way (1) from room 4.02 to room 4.99, (2) from room 4.99 to room 2.01, and (3) from room 2.01 to room 4.64. Secondly, when participants reach the destination of the third assignment, an evacuation alarm is activated and participants need to evacuate from room 4.64 to any exit on the first floor. The assignments are constructed in this way so that the complexity level increases as the variation of the assignment changes.

### 3.2 Experiment setup

Two types of VR setups were adopted in the experiment, namely a Head-mounted-display (HMD) VR and a Desktop VR. The HMD VR setup consists of an HTC Vive device (resolution of 1080 × 1200 pixels per eye, a 110-degree field of view, a 90 Hz refresh rate), a wireless controller, and a headphone. Participants used the controller to move through the virtual environment and used head orientation to control the direction of the movement. The Desktop VR setup consists of an AOC G2460PF desktop monitor (24-inch, resolution of 1920 × 1080 pixels, a 144 Hz refresh rate, 1 ms response time), a mouse, a keyboard, and a headphone. Participants rotated the mouse to change the direction of view and pressed the keyboard key 'w' to move forward in the virtual environment.

Besides the difference in technical setups, there were no other differences regarding the virtual environment, viewpoint, movement speed, etc. In both setups, software including Unreal Engine and SteamVR were used to run WayR on a desktop (AMD Ryzen 7 2700X with a 3.7 GHz CPU, MSI NVIDIA GeForce RTX 2080 graphics card, 16 GB system memory, and a Samsung 970 EVO MZ-V7E500BW 500 GB SSD). In the virtual environment, the participant's viewpoint is represented by a camera in which the participant can have a first-person perspective. Based on the pilot study, the walking speed of participants was set at 1.4 m/s to avoid simulation sickness, which is also consistent with the literature (Choi et al., 2014; Fitzpatrick et al., 2006).

### 3.3 Experiment procedure

The experiment includes four parts, namely introduction, familiarization, official VR experiment, and post-questionnaire. Firstly, participants were briefly introduced to the experiments with written instructions about the usage of devices, the general procedure of the experiment, and safety measures. Secondly, participants were asked to wear the headset or sit in front of the desktop to perform a familiarization session. Participants needed to find their way in a simple virtual scenario to get used to the devices and navigation in a virtual environment. Thirdly, once participants confirmed that they felt confident and comfortable with the devices, the official VR experiments started. Participants were asked to perform the four wayfinding assignments as described above. Once participants finished the last evacuation assignment (i.e., reaching an exit on the ground floor), the official VR experiment ended. Lastly, participants were asked to fill in a post-questionnaire in the same room.

### 3.4 Data collection process

Two types of data were collected during the experiment, namely participant's movement in the virtual building and the questionnaire data. Participants' movement trajectories were automatically recorded at a frequency of 10 Hz via Unreal Engine. The recorded movement trajectory data can be translated into wayfinding behavior regarding (1) route and exit choices, (2) wayfinding performance (i.e., time, speed, and distance), and (3) observation behavior (e.g., hesitation, heard rotation). The questionnaire data included three parts, namely (1) personal information, (2) user experience, and (3) wayfinding style. The personal information part collected participants' socio-demographic information and their experience with VR, computer gaming, and the faculty building. The user experience part includes the face validity questionnaire (Kaptein et al., 1996), the Simulator Sickness Questionnaire (Kennedy et al., 1993), the System Usability Scale (Brooke, 1996), and the Presence Questionnaire (Witmer et al., 2005). They measured the level of realism of the virtual environment, simulation sickness, usability of the VR system, and sense of presence respectively. The last part of the questionnaire measured participants' attitudes toward spatial knowledge and orientation attitude during wayfinding (Zomer et al., 2019). Participants needed to rate the following statements from 1 (strongly disagree) to 5 (strongly agree), including 'I generally choose the route that is the quickest in time', 'I generally choose the route that is shortest in distance', 'I generally choose the route that is the most direct', 'I generally choose the route that requires the least changes of direction', 'My sense of orientation is bad', and 'I often vary my route when traveling towards a known destination'.

### 3.5 Participant characterization

In total, 72 participants were recruited for the experiment via advertisements at Delft University of Technology. No one was informed about the purpose of the experiment or the layout of the virtual building. Participants received no compensation for taking part in the experiment. Two participants from the HMD VR setup took a break and did not finish the experiment. In the end, the data of 70 participants were included in the analysis. The descriptive information of the participants is presented in Table 1. 70 participants were taken into account in the modeling process. 36 of them used an HMD VR system and 34 participants used a Desktop VR. The gender distribution of the participants was relatively balanced. Most of them were familiar with the building (55 participants), had a Master's degree or higher (49 participants), had rare experience with VR systems (51 participants), and had moderate to very familiar experience with computer gaming (49 participants). Please note, as the experiment was performed at the university, participants with a BSc degree are current MSc students, participants with an MSc degree are current Ph.D. students, and participants with a doctoral degree are staff members.

Table 1. The Descriptive information of the participants.

| Descriptive information | Category | Number (percentage) |
|---|---|---|
| VR setup | HMD VR | 36(51.43%) |
| | Desktop VR | 34(48.57%) |
| Gender | Male | 41 (58.57%) |
| | Female | 29 (41.43%) |
| Familiarity with the faculty building | Not at all familiar | 0 ( 0.00%) |
| | A-little familiar | 6 ( 8.57%) |
| | Moderately familiar | 9 (12.86%) |
| | Quite-a-bit familiar | 16 (22.86%) |
| | Very familiar | 39 (55.71% |
| Highest education level | High school or equivalent | 5 (7.14%) |
| | Bachelor's degree or equivalent | 16 (22.86%) |
| | Master's degree or equivalent | 40 (57.14%) |
| | Doctoral degree or equivalent | 9 (12.86%) |
| Previous experience with VR | Never | 18 (25.71%) |
| | Seldom | 33 (47.14%) |
| | Sometimes | 15 (21.43%) |
| | Often | 1 ( 1.43%) |
| | Very often | 3 ( 4.29%) |
| Familiarity with any computer gaming | Not at all familiar | 9 (12.86%) |
| | A-little familiar | 12 (17.14%) |
| | Moderately familiar | 13 (18.57%) |
| | Quite-a-bit familiar | 14 (20.00%) |
| | Very familiar | 22 (31.43%) |
| My sense of orientation is bad | Totally disagree | 33 (47.14%) |
| | Disagree | 18 (25.71%) |
| | Natural | 10 (14.29%) |
| | Agree | 7 (10.00%) |
| | Totally agree | 2 ( 2.86%) |

## 4. MODELING STRATEGY

This study develops four distinct models, being a route choice model, a wayfinding performance model, a gazing model, and a hesitation model. The methodology to estimate a route choice model is presented in section 4.1. Accordingly, section 4.2 presents the methodology to estimate the three other models, as the same model structure and search procedure is adopted for all three. Next, section 4.3 presents the independent variables that are introduced in the modeling process. The last section presents the general trends in the distribution of the input variables.

### 4.1 Modeling route choice behavior

As identified in section 2, there is little research featuring pedestrian route choice behavior in buildings. For this preliminary study, we choose to use the two simplest route choice models that are often adopted for outdoor route choice behavior, namely the Multinomial Logit model (MNL) and the Path Size Logit model (PSL). We have explicitly opted to include both models because we question whether the overlap between routes has a similar impact on pedestrian route choice with level changes inside of buildings as it would have in outdoor environments.

#### 4.1.1 Model specification

The MNL model is defined by eq. 1 and 2. Here, $r$ represents route, $\beta_c$ and $\beta_{c,p}$ are the model weights of the route-specific characteristics $c$ and person-specific characteristics $p$ , $X_c(r)$ and $X_p$ are the route and person-specific variables respectively.

$$P(r) = \frac{e^{U(r)}}{\sum_{r \in R} e^{U(r)}} \qquad (1)$$

$$U(r) = \sum_{c \in C} \beta_c \cdot X_c(r) + \sum_{c \in C} \sum_{p \in P} \beta_{c,p} \cdot X_p \cdot X_c(r) \quad (2)$$

In the PSL model, one additional variable is added to the utility function $U(r)$ formulation (see eq. 3 and 4). This overlap factor $PS_r$ accounts for the overlap between route $r$ and all other routes in the route set $R$ . $w_l$ is the lengths of link $l$ , and $w_{r,tot}$ is the total length of route *r*. When route *r* overlaps a lot with other routes, $PS_r$ goes to 0 and the logarithm of $PS_r$ goes to negative infinity.

$$U(r) = \sum_{c \in C} \beta_c \cdot X_c(r) + \sum_{c \in C} \sum_{p \in P} \beta_{c,p} \cdot X_p \cdot X_c(r) + \beta_{PS_r} \cdot \log(PS_r) \quad (3)$$

$$PS_r = \sum_{l \in L} \frac{w_l}{w_{r,tot}} \cdot \frac{1}{\sum_{r \in R} \delta_{r,l}} \quad (4)$$

#### 4.1.2 Route set development

The input of the route choice model is a route set. In this study, the route set is developed in 3 stages. In the first stage, a breadth-first search on link elimination (BFS-LE) algorithm is adopted to create a comprehensive route set (Rieser-Schüssler et al., 2013). This algorithm was developed for high-density networks. The idea behind this approach is to calculate the shortest path, add this path to the choice set, and accordingly systematically remove each of the links of this shortest path. In each step, a new shortest path is calculated and added to the choice set. In this study, the travel distance and Dijkstra's algorithm (Dijkstra, 1959) are used to calculate the shortest path. We applied a tree-depth of two. Given the relative simplicity of the network, this results in approximately 100 distinct routes for each task. Accordingly, in stage 2, a random 30 routes are selected from the comprehensive set. Even though this limits the diversity of the route set, this number of routes is still a lot higher than the seven different alternative routes that participants on average recall in Hoogendoorn-Lanser (2005). Finally, a last check is performed. If the routes that the participants adopted are not part of the selected 30 routes, the 30th route is swapped with the chosen route.

#### 4.1.3 Search strategy

In total, four distinct discrete models are estimated, namely two MNL and two PSL models. For each type of model, one model only featuring route and infrastructure characteristics, and another model featuring route, infrastructure, and personal characteristics is estimated. A stepwise combinatory scheme is adopted to search for the best model. For all model structures, the same procedure is followed. In stage 1, a series of models is estimated featuring one variable at a time. Accordingly, only models that feature significant variables are taken to the next stage. In stage 2, we add one additional variable to all significant models from stage 1. All combinations are tested again. Only models that show significant improvements concerning the single-variable models and feature only significant variables continue to stage 3, where again one additional variable is added. The best infrastructure model is used as a start for the development of the comprehensive model featuring both infrastructure and personal characteristics. See Figure 2 for a visualization of the search strategy.

A t-test ($\alpha = 95\%$) is adopted to test the significance of the weights $\beta$ in each model. A $\chi^2$ test is adopted to test for a significant improvement between model stages. Here, in each following stage, the new model extension is tested against all other models that feature a subset of the model's variables. For example, F (A, B) is tested against F(A) and F(B). At the moment that no more variables can be added or no significant improvements are possible anymore, the final models are compared using AIC and BIC, because the largest models with the same amount of variables are per definition no subsets of each other.

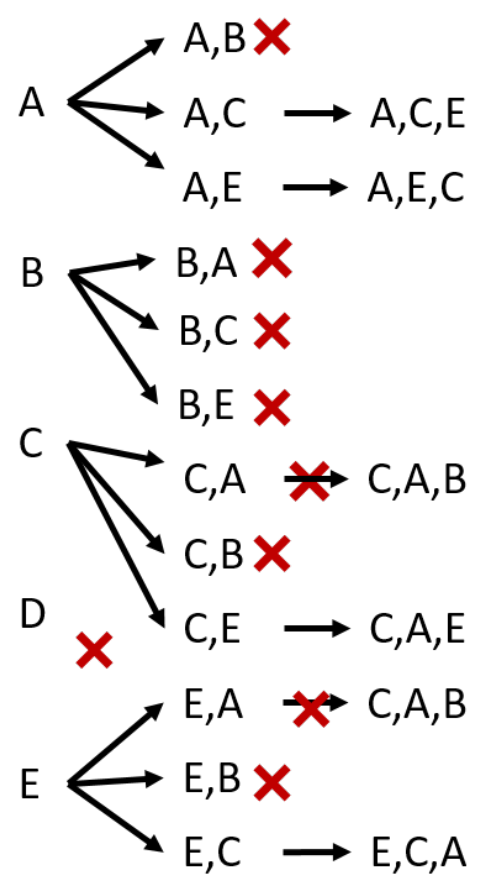

Figure 2. Visualization of search strategy (x means non-significance)

### 4.2 Modeling wayfinding performance and observation behavior

The literature featuring wayfinding performance and observation behavior (i.e., hesitation and gazing behavior) presents predominantly one-to-one relations between independent variables and one specific behavioral metric. In this paper, we are especially interested in the combinatory impact of multiple variables (i.e., personal, route, and infrastructure characteristics) on pedestrian wayfinding performance and observation behavior. Here, wayfinding performance is operationalized as the total travel time per task. Regarding observation behavior, hesitation behavior is operationalized as the total number of hesitations per task, where one hesitation is a pause for longer than 3 seconds (Suzer et al., 2018). And gazing behavior is operationalized as the average head rotation change.

#### 4.2.1 Model specification

The output variables of all three models are continuous. The one-to-one relations between the input variables and output variables are predominantly linear relationships. As the number of records is limited, advanced machine-learning approaches cannot be used. Instead, a Multiple Linear Regression model (MLR) is adopted, see eq. 5 for the model formulation. Here, the $\beta_0$ represents the intercept, $\beta_1$ till $\beta_n$ represent the parameters of the model, $X_1$ till $X_n$ represent the independent variables that are taken into account in a particular model, and $\varepsilon$ represents the error term.

$$v(i) = \beta_0 + \beta_1 X_1 + \beta_2 X_2 + \cdots + \beta_n X_n + \varepsilon \tag{5}$$

#### 4.2.2 Search strategy

Three distinct MLR models are estimated per pedestrian behavior (i.e., walking performance, hesitation, and gazing). The first MLR model only features variables that describe the characteristics of chosen route, infrastructure, and task. The second MLR model incorporates the personal characteristics of participants. The third MLR model combines route, infrastructure, task, and personal characteristics.

For all nine models (i.e., 3 x 3), a similar search strategy is applied. All models are estimated using IBM SPSS Statistics 28.0 (IBM Corp) featuring a backward stepwise search approach. This approach allows us to account for collinearity between the variables. In a backward stepwise search, first, a model containing all variables is estimated. Afterward, the least significant variables are taken out of the full model one after another. This process continues until no other variables can be taken out without decreasing the explainability of the model significantly. The F-test is used to determine which variables are taken out of the full model next, with an entry and removal probability of 0.05.

### 4.3 Introduction independent variables

Pedestrians are influenced by a large array of factors while determining their route and walking speed. This study focuses on the factors that are related to the person, the infrastructure they walk through, the route that they choose, and the wayfinding task they need to perform. Below we provide an in-depth introduction to the variables in each category. In Table 2, an overview of all independent variables is provided, including their main descriptive.

### 1. Participant characteristics

Eight personal characteristics of participants were recorded during the VR study, including age, gender, education level, height, familiarity with the building, orientation ability, gaming experience, and VR experience. A detailed explanation of these personal characteristics is provided below.

For the first factor, ***age***, we expect a non-linear effect for route choice as most young participants (i.e., students) have a different activity pattern in this building. Literature shows the walking speed of younger individuals tends to be highest at the age of approximately 20-25 years (Weidman, 1993). To account for this non-linearity, we introduce the continuous variable age, as well as two binary indicator variables for young (age_young: <25 years old) and old (age_old: >50 years old) participants. For ***gender***, one binary indicator variable (1 = male) is created, as no non-binary gender responses were reported by the participants. ***Education level*** is translated into a set of four binary indicator variables based on the highest achieved level of education, including Secondary Education, Bachelor's degree, Master's degree, and Postdoctoral degree. Please note, the indicator variable for secondary education also features participants that finalized a practical college degree (VMBO, MBO) in The Netherlands. Moreover, in this study, most participants with a finished secondary education degree are current BSc students, and most participants with a finished BSc degree are current MSc students. Similarly, participants with a finished MSc degree are predominantly PhD students, and participants with a finished PhD degree are staff members. For ***height***, participants were asked to fill in their height in the questionnaire.

The variable ***familiarity with the building*** was a 5-point Likert-scale question. This response is translated into two binary indicator variables for those that are familiar ('quite a bit familiar' or 'very familiar') and those that are not familiar ('not at all familiar', 'a little familiar' or 'moderately familiar'). The variable, ***orientation***, indicates how the participants score their orientation ability. This question is translated into two binary variables for those with good self-reported orientation abilities ('disagree' or 'totally disagree') and bad self-reported orientation abilities ('neutral', 'agree' or 'totally agree').

To account for their familiarity with VR systems, we have asked the respondents two questions. The first relates to general familiarity with gaming, here coined ***gaming experience.*** This 5-point Likert-scale question is translated into two binary indicator variables, one featuring frequent gaming experience ('quite a bit familiar' or 'very familiar') and one featuring limited gaming experience ('moderately familiar', 'a-little familiar', or 'not at all familiar'). The last 3-point Likert scale question featured their ***experience with VR systems.*** Here, three binary indicator variables are coded, featuring high VR familiarity ('often' or 'very often'), medium VR familiarity ('sometimes' or 'seldom'), and no VR familiarity ('never').

### 2. Infrastructure variables

The building has five distinguishing features. First, only one side of the two long corridors features windows. Therefore, we account for the ***percentage of distance covered on links with a window (window)*** that the participants passed during each assignment. Second, ***fire doors (firedoor)*** are located in both parallel corridors. In the VR environment these doors are always open, yet the cross-section is narrower. Third, we account for the ***number of floor signs*** (***floorsigns***) that a certain route passed. Fourth, the building consists of multiple ***floor levels*** (***level_no***). We account for the number of floors that were visited by the participant during an assignment. If participants visit a floor more than once, we account for them double (i.e., 3 – 2 – 1 – 2 = 4 floor levels). The same goes for the usage of staircases (***stairs_no***).

### 3. Route variables

The participant's route choice is coded into nine distinct variables. We use figure 3 to explain each of the variables. The ***total length of the route (distot***) (i.e., [5 + 10 + 12 + 10 + 25 + 10 + 2] = 74 m) is coded as a continuous variable. Other continuous variables include the ***distance to the first turn*** (***dist_firstturn***) (i.e., 5 m), the ***average length of all straight stretches (dist_avg_straight)*** of the route (i.e., [5 + 10 + 12 + 10 + 25 + 10 + 2] / 7 = 10.58 m), and the ***longest straight stretch (dist_longeststretch)*** of the route (i.e., 25m).

To describe the turning behavior of the participants, we have adopted three distinct variables representing the ***number of turns*** a participant makes (i.e., **turns_tot**, **turns_left**, **turns_right**). Here, we account for a turn if the participant moves towards a new link that has, at least, a 90-degree angle with respect to the previous link. In the example of figure 3, the total number of turns is 6, of which there are three turns to the left and three turns to the right. Because turns can be sharp or narrow, we also account for the ***total amount of turning*** (***rot_abs***) that participants undertook. This variable sums the absolute angles in degrees of all turns a participant made (i.e., 540 degrees in the example). The rotations that pedestrians make as part of level changes are not accounted for in this number. The last route variable, ***ratio_wide,*** accounts for the time participants spend in the wider parallel corridors. This variable is operationalized as the total travel distance in the wider corridors divided by the total length of the route that participants traveled.

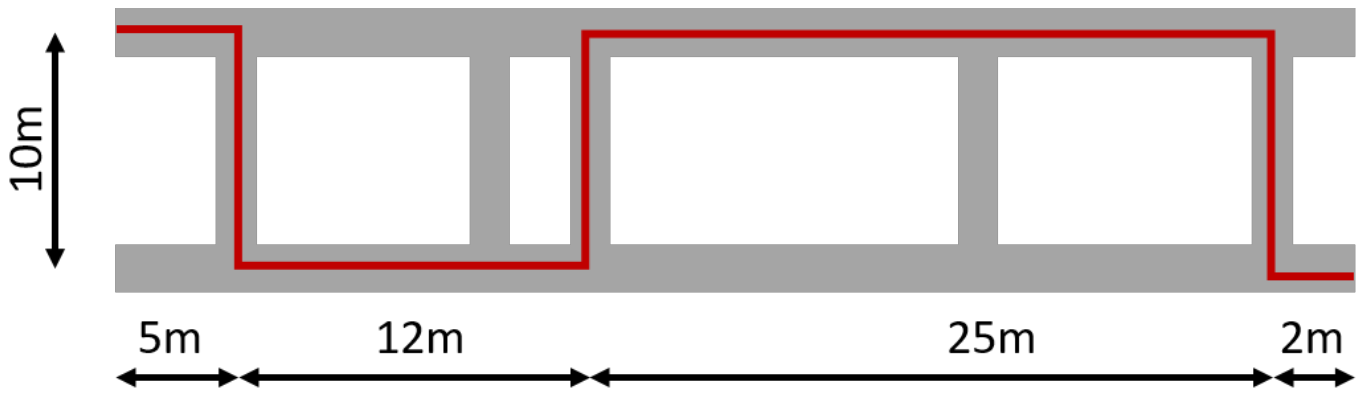

Figure 3. Example route to explain route variables.

**4. Task variable**

As mentioned in section 3, each participant performs four wayfinding ***tasks***. As a result, four binary indicator variables were adopted to identify which of the task is performed (e.g., task_1). This indicator variable is included to account for any potential deviations in behaviors related to the complexity of the wayfinding assignments.

Table 2. The variables introduced during the modeling process.

| **Variable** | **Type** | **Min** | **Max** | **Median** | **Std** | **Explanation** |
|---|---|---|---|---|---|---|
| Age | Integer | 17 | 64 | 28 | 6.35 | Age in years |
| Age_young | binary | 0 | 1 | 0 | 0.46 | Dummy for all people aged 25 or younger |
| Age_old | binary | 0 | 1 | 0 | 0.48 | Dummy for all people aged 50 or older |
| Gender | binary | 0 | 1 | 1. | 0.44 | Binary (male = 1) |
| Education_Sec | binary | 0 | 1 | 0 | 0.12 | Highest level of education is a secondary school degree |
| Education_BSc | binary | | 1 | 1 | 0.49 | Highest level of education is a Bachelor degree |
| Education_MSc | binary | 0 | 1 | 0 | 0.26 | Highest level of education is a Master degree |
| Education_Doc | binary | 0 | 1 | 0 | 0.42 | Highest level of education is a PhD degree |
| Height | continuous | 151 | 196 | 175 | 9.28 | Height of the person |
| Familiar | binary | 0 | 1 | 1 | 0.50 | Person is familiar with the building |
| Familiar_not | binary | 0 | 1 | 0 | 0.41 | Person is not familiar with the building |
| Gaming_often | binary | 0 | 1 | 0 | 0.34 | Person plays video games often |
| Gaming_not | binary | 0 | 1 | 0 | 0.50 | Person does not play video games often |
| VR_often | binary | 0 | 1 | 1 | 0.41 | Person has often experienced VR |
| VR_sometimes | binary | 0 | 1 | 1 | 0.50 | Person has sometimes experienced VR |
| VR_never | binary | 0 | 1 | 0 | 0.12 | Person has never experienced VR |
| Orientation_good | binary | 0 | 1 | 1 | 0.47 | Person's orientation ability is good |
| Orientation_bad | binary | 0 | 1 | 1 | 0.48 | Person's orientation ability is bad |
| Window | integer | 0 | 0.19 | 0.04 | 0.05 | Number of windows on the route |
| Firedoor | integer | 0 | 5 | 4 | 1.54 | Number of fire doors on route |
| Floorsigns | integer | 1 | 6 | 4 | 1.26 | Number of floor signs on the route |
| Level_no | integer | 0 | 4 | 1.24 | 3 | Number of level changes |
| Stairs_no | integer | 0 | 4 | 1.11 | 2 | Number of stairs traveled |
| Distot | continuous | 4144 | 23206 | 18824 | 6486 | Total distance traveled (cm) |
| Dist_firstturn | continuous | 0 | 20158 | 2070 | 4263 | Distance until first turn (cm) |
| Dist_avg_straight | continuous | 255 | 10799 | 2655 | 3361 | Distance of the average straight stretch (cm) |
| Dist_longeststretch | continuous | 1530 | 20158 | 10800 | 5894 | Distance of the longest straight stretch (cm) |
| Turns_tot | integer | 0 | 8 | 1.62 | 2 | Number of turns |
| Turns_left | integer | 0 | 3 | 0.70 | 0 | Number of turns to the left |
| Turns_right | integer | 0 | 3 | 0.69 | 0 | Number of turns to the right |
| Rot_abs | continuous | 0 | 1080 | 180 | 180 | Sum of the absolute rotations (◦) |
| Ratiowide | continuous | 0.37 | 0.97 | 0.89 | 0.22 | Percentage of distance traveled on wide corridors (cm) |
| Task_1 | binary | 0 | 1 | 0 | 0.43 | Dummy for task 1 |
| Task_2 | binary | 0 | 1 | 0 | 0.43 | Dummy for task 2 |
| Task_3 | binary | 0 | 1 | 0 | 0.43 | Dummy for task 3 |
| Task_4 | binary | 0 | 1 | 0 | 0.43 | Dummy for task 4 |

### 4.4 General statistics & correlations between variables

Two correlation analyses are performed to identify correlations between the independent variables (see Tables 3 and 4). The results show that the infrastructure variables are highly correlated. Thus, we expect that only a subset of infrastructure variables will be included in the final models. In contrast, little correlations are found between different personal characteristics. Most interestingly, a very strong negative (-1.0) correlation exists between VR experience and familiarity with the building. We expect that this is due to the limited number of participants that have this combination of characteristics. Similarly, we see a strong negative correlation between age and gender, identifying that a large part of our 'older' participants is female. Given the limited number of strong correlations, most personal characteristics can be included in the modeling process simultaneously.

Table 3. Spearman's rank correlation coefficient (ρ) for the infrastructure variables, where the fill of the cell indicates a strong negative (red) or positive (green) correlation larger than 0.4.

| | distot | dist_firstturn | dist_longeststretch | dist_avg_straight | ratiowide | level_no | turns_tot | turns_left | turns_right | rot_abs | window | firedoor | floorsigns | task_1 | task_2 | task_3 |
|---|---|---|---|---|---|---|---|---|---|---|---|---|---|---|---|---|
| dist_firstturn | -0.15 | | | | | | | | | | | | | | | |
| dist_longeststretch | 0.81 | -0.26 | | | | | | | | | | | | | | |
| dist_avg_straight | 0.77 | 0.05 | 0.83 | | | | | | | | | | | | | |
| ratiowide | 0.70 | 0.09 | 0.81 | 0.98 | | | | | | | | | | | | |
| level_no | -0.61 | -0.10 | -0.69 | -0.92 | -0.93 | | | | | | | | | | | |
| turns_tot | 0.85 | -0.13 | 0.60 | 0.65 | 0.53 | -0.49 | | | | | | | | | | |
| turns_left | 0.60 | -0.06 | 0.37 | 0.59 | 0.50 | -0.62 | 0.78 | | | | | | | | | |
| turns_right | 0.58 | -0.09 | 0.39 | 0.59 | 0.49 | -0.63 | 0.74 | 0.88 | | | | | | | | |
| rot_abs | 0.80 | -0.17 | 0.54 | 0.47 | 0.35 | -0.23 | 0.93 | 0.55 | 0.50 | | | | | | | |
| windows | 0.32 | 0.09 | 0.27 | 0.41 | 0.44 | -0.55 | 0.18 | 0.48 | 0.48 | -0.02 | | | | | | |
| firedoor | 0.88 | -0.10 | 0.83 | 0.90 | 0.85 | -0.81 | 0.73 | 0.64 | 0.63 | 0.57 | 0.38 | | | | | |
| floorsigns | 0.92 | -0.10 | 0.85 | 0.91 | 0.85 | -0.80 | 0.78 | 0.63 | 0.65 | 0.64 | 0.37 | 0.97 | | | | |
| task_1 | 0.26 | 0.19 | 0.40 | 0.76 | 0.76 | -0.81 | 0.34 | 0.55 | 0.56 | 0.08 | 0.31 | 0.55 | 0.52 | | | |
| task_2 | 0.75 | -0.36 | 0.50 | 0.25 | 0.16 | 0.01 | 0.62 | 0.21 | 0.17 | 0.77 | -0.05 | 0.50 | 0.52 | -0.33 | | |
| task_3 | -0.25 | 0.14 | -0.15 | -0.25 | -0.15 | 0.01 | -0.51 | -0.30 | -0.26 | -0.56 | 0.36 | -0.25 | -0.23 | -0.33 | -0.33 | |
| task_4 | -0.76 | 0.03 | -0.75 | -0.76 | -0.77 | 0.80 | -0.45 | -0.46 | -0.47 | -0.29 | -0.61 | -0.80 | -0.81 | -0.33 | -0.33 | -0.33 |

Table 4. Spearman's rank correlation coefficient (ρ) for the socio-demographic variables, where the fill of the cell indicates a strong negative (red) or positive (green) correlation larger than 0.4.

| | age | age_young | age_old | gender | height | education_Sec | education_BSc | education_MSc | education_Doc | familiar | familiar_not | gaming_often | gaming_not | VR_often | VR_sometimes | VR_never | orientation_good |
|---|---|---|---|---|---|---|---|---|---|---|---|---|---|---|---|---|---|
| age_young | 0.00 | | | | | | | | | | | | | | | | |
| age_old | -0.06 | -0.05 | | | | | | | | | | | | | | | |
| gender | -0.76 | -0.04 | -0.16 | | | | | | | | | | | | | | |
| height | -0.15 | 0.06 | 0.18 | 0.16 | | | | | | | | | | | | | |
| education_Sec | 0.21 | 0.18 | -0.09 | -0.20 | 0.01 | | | | | | | | | | | | |
| education_BSc | -0.02 | -0.15 | -0.01 | 0.10 | 0.56 | 0.10 | | | | | | | | | | | |
| education_MSc | -0.33 | -0.06 | 0.25 | 0.16 | 0.31 | -0.03 | 0.12 | | | | | | | | | | |
| education_Doc | -0.47 | 0.01 | 0.00 | 0.32 | -0.20 | -0.07 | -0.09 | -0.15 | | | | | | | | | |
| familiar | 0.27 | 0.00 | -0.11 | -0.05 | 0.05 | -0.14 | -0.03 | -0.32 | -0.63 | | | | | | | | |
| familiar_not | -0.29 | 0.11 | -0.04 | 0.23 | 0.06 | -0.06 | -0.06 | 0.13 | 0.30 | -0.18 | | | | | | | |
| gaming_often | 0.45 | 0.03 | -0.03 | -0.46 | -0.08 | 0.31 | 0.06 | -0.11 | -0.21 | -0.44 | -0.20 | | | | | | |
| gaming_not | 0.05 | 0.11 | -0.10 | -0.15 | -0.29 | 0.12 | -0.63 | -0.16 | 0.15 | -0.02 | 0.12 | -0.03 | | | | | |
| VR_often | 0.29 | -0.11 | 0.04 | -0.23 | -0.06 | 0.06 | 0.06 | -0.13 | -0.30 | 0.18 | -1.00 | 0.20 | -0.12 | | | | |
| VR_sometimes | -0.05 | -0.11 | 0.10 | 0.15 | 0.29 | -0.12 | 0.63 | 0.16 | -0.15 | 0.02 | -0.12 | 0.03 | -1.00 | 0.12 | | | |
| VR_never | -0.03 | -0.08 | 0.16 | 0.07 | 0.16 | -0.01 | 0.10 | -0.03 | -0.07 | 0.10 | -0.06 | -0.05 | -0.12 | 0.06 | 0.12 | | |
| orientation_good | 0.01 | -0.97 | 0.01 | 0.02 | -0.10 | -0.18 | 0.12 | 0.07 | 0.00 | -0.03 | -0.10 | -0.02 | -0.08 | 0.10 | 0.08 | -0.18 | |
| orientation_bad | 0.06 | 0.05 | -1.00 | 0.16 | -0.18 | 0.09 | 0.01 | -0.25 | 0.00 | 0.11 | 0.04 | 0.03 | 0.10 | -0.04 | -0.10 | -0.16 | -0.01 |

## 5. RESULTS

This section presents the results of the models featuring pedestrian wayfinding behavior. Section 5.1 presents the results of route choice models. Section 5.2 presents the results of models featuring wayfinding performance. And the last section presents the modeling results featuring observation behavior.

### 5.1 Modelling route choice

In this section, the generated routes are first presented. Accordingly, section 5.1.2 presents the route choice models. Section 5.1.3 compares our findings to route choice model results presented by other studies.

#### 5.1.1 Generated route set

Figure 4 provides a visualization of the spatial distribution of the generated routes across the network for two of the four assignments. As one can see, the generated routes are well distributed. That is, an equal number of routes makes use of each side of the corridor in assignment 1, and there is an even distribution concerning the stairs that are adopted to travel from floor 4 to floor 2 in assignment 2. Moreover, the statistics in Table 5 show that the variability in the dataset is fairly similar to the variability in the adopted routes. The median and standard deviation of most infrastructure and route characteristics are of similar proportions.

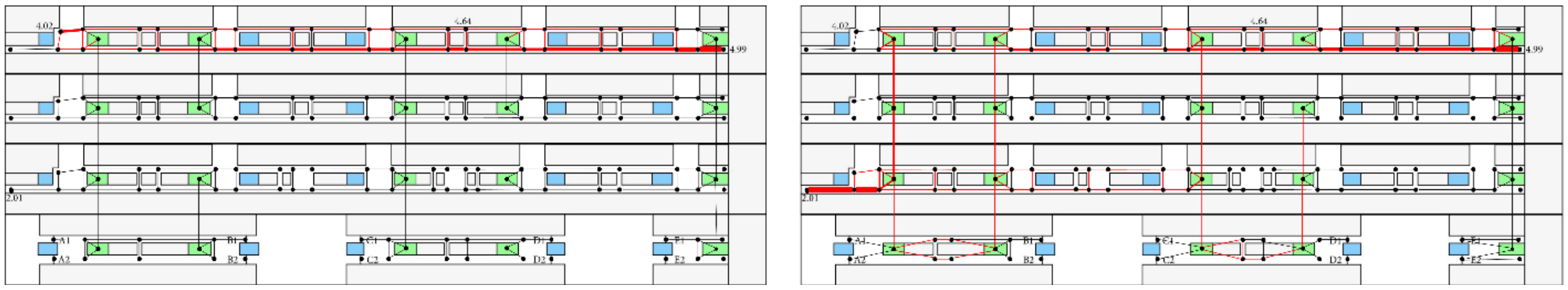

Figure 4. Spatial distribution of generated route sets through the building for assignments 1 and 2

Table 5. Overview of statistics for the generated routes.

| | | Route set | | | | Chosen Route | |
|---|---|---|---|---|---|---|---|
| | | Min | Max | Median | Std. | Median | Std. |
| Infrastructure characteristics | Windows | 0 | 0.19 | 0.03 | 0.04 | 0.04 | 0.05 |
| | Firedoor | 0 | 5 | 3 | 1.43 | 4 | 1.54 |
| | Floorsign | 1 | 6 | 3 | 1.19 | 4 | 1.26 |
| | Floor levels | 0 | 4 | 2 | 1.20 | 1.24 | 3 |
| Route characteristics | Disttot | 41.44 | 232.06 | 188.24 | 52.04 | 188.24 | 64.86 |
| | Dist_longeststretch | 15.30 | 201.58 | 94.04 | 44.83 | 108.00 | 58.94 |
| | Dist_avg_straight | 2.40 | 107.99 | 26.55 | 21.32 | 26.55 | 30.61 |
| | Dist_firstturn | 0 | 201.58 | 22.04 | 48.02 | 20.70 | 42.63 |
| | Turns_tot | 0 | 8 | 4 | 1.64 | 1.62 | 2 |
| | Turns_left | 0 | 4 | 1 | 1.06 | 0.70 | 0 |
| | Turns_right | 0 | 3 | 1 | 1.07 | 0.69 | 0 |
| | Rot_abs | 0 | 1080 | 450 | 195.00 | 180 | 180 |
| | Ratio_wide | 0.31 | 0.97 | 0.86 | 0.14 | 0.89 | 0.22 |
| | Overlap factor | 0.023 | 0.664 | 0.106 | 0.076 | n.a. | n.a. |

**5.1.2 Route choice model results**

Using generated route sets, four distinct route choice models were estimated by means of the modeling methodology presented in section 4.1. In particular, two MNL and PSL models are developed. One only features infrastructure and route characteristics. The other also includes personal characteristics. Table 6 presents these four choice models.

Adding interaction variables to the model slightly improves the model fit in both MNL and PSL models. The hypotheses that the improved models (those with personal characteristics) are the better models are accepted for the MNL model ($\chi2$ (3, $N$ = 280) = 20.58, $p$ <.01) as well as the PSL model ($\chi2$ (2, $N$ = 280) = 9.18, $p$ <.01). Compared to the MNL models, the PSL models better explain the variability. The model with the best fit is the PSL model featuring both infrastructure and personal characteristics. Of the two PSL models, which are not nested versions, the PSL infra model[1] is the best (i.e., BIC 700.4 < 702.5 < 791.6 < 795.3). This shows that the actual improvement in the ability of the comprehensive models (i.e., infrastructure and personal characteristics) to explain the variability of the data is limited ($\Delta\rho_{adj}$ = 0.002). This result indicates that personal characteristics have little explanatory power regarding route choice. Whether this is due to the limited number of participants, limited variability of the personal characteristics in the data set, or the fact that personal characteristics are not significantly influencing route choice is difficult to identify.

The PSL infra model shows that pedestrians prefer routes that have a lot of overlap with other routes in the route set. Moreover, three route characteristics are featured. Pedestrians prefer shorter routes, long straight stretches, and stretches where the average length per stretch is relatively large. Meanwhile, three infrastructure characteristics are featured in the final model, namely the number of floor levels visited, the number of floor signs along the route, and the percentage of a route that features windows. Surprisingly, all three parameters are positive. We would expect a negative parameter for the number of levels, as we expect pedestrians to prefer the lowest number of level changes along their route. Potentially, the number of levels is an offset against the overlap factor, as many direct routes are part of the route set.

[1] For the sake of simplicity, all the models feature infrastructure and route characteristics are referred to as the infra model.

Table 6. Four different route choice models for pedestrian route choice in multi-story buildings.

| | **MNL infra** | | **MNL infra + personal char** | | **PSL infra*** | | **PSL infra + personal char** | |
|---|---|---|---|---|---|---|---|---|
| | **Beta** | **p-value** | **Beta** | **p-value** | **Beta** | **p-value** | **Beta** | **p-value** |
| Dist_tot | 0.36 | 0.00 | -0.225 | <0.01 | -0.216 | <0.01 | -0.216 | <0.01 |
| Dist_firstturn | -0.013 | 0.02 | -0.0165 | <0.01 | | | | |
| Dist_avg_straight | -0.016 | 0.00 | 0.28 | <0.01 | 0.126 | 0.02 | 0.191 | <0.01 |
| x age_young | | | 0.239 | 0.03 | | | | |
| x edu_sec | | | 0.4412 | <0.01 | | | | |
| x age_old | | | | | | | -2.15 | 0.03 |
| Dist_longeststretch | -0.218 | 0.00 | 0.0208 | <0.01 | 0.018 | <0.01 | 0.017 | <0.01 |
| x age_young | | | -0.0221 | 0.01 | | | | |
| Number of levels | | | | | 3.52 | <0.01 | 3.61 | <0.01 |
| Floorsigns | | | | | 3.63 | <0.01 | 3.60 | <0.01 |
| Window | | | | | 14.9 | <0.01 | 13.9 | <0.01 |
| x education MSc | | | | | | | 13.4 | 0.03 |
| Overlap factor | | | | | -29.9 | <0.01 | -30 | <0.01 |
| Log-likelihood | -386.38 | | -376.10 | | -330.48 | | -325.89 | |
| Rho2 | 0.594 | | 0.598 | | 0.646 | | 0.648 | |
| AIC | 780.8 | | 766.2 | | 674.9 | | 669.8 | |
| BIC | 795.3 | | 791.6 | | 700.4 | | 702.5 | |

* Best fit model

### 5.1.3 Discussion of the route choice results

Our findings have some resemblance with the literature featuring pedestrian route choice behavior. First, we find the overall distance of the route has a negative impact on the probability that a route is adopted. This finding is consistent with, amongst others, the studies of Borgers and Timmermans (1986) and Stubenschrott et al. (2014). Second, participants prefer routes with relatively long stretches, which often feature a low number of turns and a limited number of decision points. Similar findings were found by (Basu et al., 2022; Dalton, 2003; Natapov and Fisher-Gewirtzman, 2016; Sevtsuk and Basu, 2022; Wiener et al., 2012), which identified that pedestrians prefer routes with the least number of turns and straighter line of vision. Third, we found that infrastructure attributes such as floor signs and routes with windows have a significant impact on pedestrian route choice. This finding is in line with studies that showed visual accessibility and guidance information are of great importance for people to find their way in the environment (Dogu and Erkip, 2000; Gärling et al., 1983; Montello and Sas, 2006; Vilar et al., 2015, 2014b). Lastly, the overlap between routes is valued by the participants in our study. This finding is in accordance with several bicycle route choice studies that found the tendency for cyclists to prefer routes that overlap heavily with other routes in the route set (e.g., Hood et al., 2011; Ton et al., 2017).

Interestingly, our study shows that the personal characteristics of participants have a limited impact on their route choice. There might be two reasons for this. First, it could be that we did not find the difference due to the fairly stable route choice of participants. In particular, the layout of the building might limit the variation in route choice for participants (i.e., two major corridors). Meanwhile, the use of the VR devices might also have influenced the route choice behavior of participants, in a way that they choose routes that are 'easier' to operate. Second, studies showed that when people are familiar with the environment, gender differences are minimal (Chen et al., 2018; Iachini et al., 2009; Nori et al., 2018; Tascón et al., 2021), which is our case that participants had a relatively high familiarity with the building.

### 5.2 Modelling wayfinding performance

Next to route choice, pedestrians also make decisions on the operational level along their route, which influences the overall wayfinding performance.  This section provides an overview of the modeling results concerning the wayfinding performance. In particular, we study the average operational walking speed of participants. First, section 5.2.1 presents the two wayfinding performance models. Section 5.2.2 compares our results to the findings of other studies.

#### 5.2.1 Wayfinding performance modeling results

In total, three distinct models have been estimated. One MLR model features infrastructure and route characteristics, one MLR model features the personal characteristics of the participants, and one MLR model features infrastructure, route, and personal characteristics. See Table 7 for the modeling results of three models.

The ***infra model*** features 7 characteristics. The average distance of the straight stretches, the average length of the straight route, the length of the longest straight stretch along the route, and the distance to the first turn have small positive effects on the wayfinding performance. Suggesting that participants prefer equally spaced straight stretches and a route that has long straight stretches. Potentially because the participants can adopt a steadier pace along the straight stretches. The number of level changes also has a positive impact on the wayfinding performance. In contrast, the percentage of windows along the route and a relatively high percentage of wide corridors have a negative impact on wayfinding performance. Surprisingly, the number of turns in the route improves wayfinding performance as well. This is unexpected, as in real-life pedestrians tend to slow down while moving around a corner (Dias et al., 2014).

The ***personal characteristics model*** features just 4 variables. This model identifies that males walk faster than females at a 90% significance level. Moreover, education level seems to have a varying impact, depending on the highest level of education received by the participants. The results show that BSc students (education_Sec) walk slower than MSc students (education_BSc). The two education variables might be a proxy for the experience of students in the faculty building. Surprisingly, participants who report bad orientation skills record a better walking performance than those that report good orientation skills.

The ***combined model*** features a large set of variables. Most variables and effects remain similar to the above-mentioned explanations regarding the infra model and personal characteristics model. In the combined model, the parameter regarding the distance until the first turn (dist_firstturn) is replaced by the parameter describing the total distance of the route (dist_tot). Here, we find the total distance of the route has a negative impact on the wayfinding performance. Two additional personal characteristics are introduced in the combined model, namely frequent gaming experience and the height of the participant. Both of them have a slightly positive impact on wayfinding performance.

Table 7. Three MLR models featuring wayfinding performance.

| | **MLR infra** | | | **MLR personal char** | | | **MLR infra + personal char*** | | |
|---|---|---|---|---|---|---|---|---|---|
| | **Beta** | **Std** | **p-value** | **Beta** | **Std** | **p-value** | **Beta** | **Std** | **p-value** |
| Constant | 1.324 | 0.021 | <0.001 | 1.091 | 0.020 | <0.001 | 0.963 | 0.182 | <0.001 |
| Dist_tot | | | | | | | 0.0015 | 0.000 | 0.008 |
| Dist_avg_straight | 0.0045 | 0.000 | <0.001 | | | | 0.0045 | 0.000 | <0.001 |
| Dist_longeststretch | 0.0011 | 0.000 | <0.001 | | | | 0.0007 | 0.000 | 0.016 |
| Dist_firstturn | 0.00036 | 0.000 | 0.037 | | | | | | |
| Ratiowide | -0.643 | 0.82 | <.001 | | | | -0.934 | 0.133 | <0.001 |
| Number of levels | 0.082 | 0.033 | 0.015 | | | | 0.073 | 0.030 | 0.016 |
| Window | -0.523 | 0.168 | 0.002 | | | | -0.369 | 0.151 | 0.016 |
| Number of turns | 0.031 | 0.004 | <0.001 | | | | 0.017 | 0.007 | 0.013 |
| Gender (male = 1) | | | | 0.037 | 0.020 | 0.061 | 0.052 | 0.015 | <0.001 |
| Education_Sec | | | | -0.161 | 0.072 | 0.027 | -0.215 | 0.051 | <0.001 |
| Education_BSc | | | | 0.061 | 0.017 | <0.001 | 0.052 | 0.014 | <0.001 |
| Orientation_bad | | | | 0.038 | 0.018 | 0.029 | 0.039 | 0.012 | 0.002 |
| Gaming_often | | | | | | | 0.045 | 0.020 | 0.025 |
| Height | | | | | | | 0.001 | 0.001 | 0.059 |
| Adj. R square | | 0.461 | | | 0.085 | | | 0.581 | |
| F stat | | 34.905 | | | 7.449 | | | 30.640 | |
| Significance | | <0.001 | | | <0.001 | | | <0.001 | |

* Best fit model

### 5.2.2 Discussion of the wayfinding performance results

There is little literature on modeling pedestrian wayfinding performance in buildings. Regarding the impact of infrastructural and route variables on wayfinding performance, firstly, we find that distractions in the environment can negatively impact wayfinding performance. In the current study, distractions are associated with loggia with different types of furniture, which are present on the side of the corridor with windows. Similar findings were also identified by Borgers and Timmermans (1986), who found a negative impact of windows in the form of shopping facades on pedestrian movement speed. Secondly, we find that wide corridors have a negative impact on wayfinding performance. That is because locations with extra information (e.g., decision points with floor plans, evacuation signs, landmarks, and room numbers) are located in the wide corridors, where participants need time to process to be able to move forward. The results are consistent with findings regarding the relationship between location and information processing during wayfinding (Conroy, 2001; Ewart and Johnson, 2021; Orellana and Al Sayed, 2013). Thirdly, our study shows a slightly positive impact of straight paths and the number of turns on wayfinding performance. Higher movement speed on a stretch path is also found in studies by Bae et al. (2021) and Zhang et al. (2011). Regarding the number of turns, the studies that present wayfinding performance models often feature the number of turns as a proxy for the complexity of a route across a flat surface. Dias et al. (2014) showed that in real life, pedestrians tend to slow down while taking a corner. Potentially, participants perform locomotion in a virtual environment plays an important role here, as they do not need to physically move in the actual environment.

Several studies suggested the impact of personal characteristics on wayfinding performance. There has been much discussion about the influence of gender differences on wayfinding, our modeling results suggest that male and taller individuals have a higher locomotion speed in the virtual building. Similar findings are identified previously by literature (Chen et al., 2009; Iachini et al., 2009). It is important to note that in the current VR study, the physical capabilities of the participants do not guide their walking speed in the virtual environment. Previous research suggests that the positive influence of gaming experience on human-computer interaction can result in better wayfinding performance in virtual environments (Castelli et al., 2008; Moffat et al., 2001; Murias et al., 2016; Smith and Du'Mont, 2009), which is also identified in our study. Moreover, the positive impact of height can be a result of the improved oversight of the scenario, as the vantage point of the participants is calibrated to their natural height. A similar impact of height is also indicated by Bosina and Weidmann (2017). Regarding the positive impact of educational level, in the current study, the education variables might be a proxy for familiarity with the building. As students (i.e., MSc students) who are more familiar with the building have better wayfinding performance. The positive impact of familiarity is also recorded in the studies of Hölscher et al. (2007) and Iachini et al. (2009). The present study does not provide a clear indication of the negative effect of highly self-rated orientation ability on wayfinding performance. It is possible that self-reported orientation ability is associated with other measures of wayfinding (e.g., spatial learning abilities) and literature suggests that self-reports often fail to capture this nuance of wayfinding ability (Castelli et al., 2008).

## 5.3 Modelling observation behavior

Two distinct models are estimated to model the participants' observation behavior. The first model explains the number of hesitations along the route (section 5.3.1). The second model captures the average head rotation of participants during wayfinding tasks (section 5.3.2).

### 5.3.1 Results and discussion of hesitation model

Three distinct hesitation models are estimated for hesitation behavior, one features infrastructure and route characteristics, one features personal characteristics, and one combines both sets of characteristics. The modeling results are presented in Table 8. All three hesitation models have a very small fit (i.e., the adjusted $R^2 < 0.25$) and introduce a large number of variables. The model that features both personal and infrastructure characteristics scores best (i.e., highest $R^2$). In addition, little overlap between the impact of the infrastructure and personal characteristics is found. Only the rot_abs and firedoor variables disappear in the combined model.

Regarding the impact of infrastructure characteristics, first, the combined model shows that turning to the right (fewer hesitations) has the opposite effect from turning to the left (more hesitations). Given the building network and the wayfinding task setting, often participants first turn to the right and then turn to the left to continue their path. It might be that turning to the right is often made in open spaces, such as loggias and intersections while turning to the left often occurs in tight corridors with little oversight of what is behind the corner. Therefore participants need more observations when truing left. Second, we find tasks 2 and 3 have a negative impact on the number of hesitations, which might be due to the learning effect as participants spend more time in the building. Third, similar to findings in the literature that suggest hesitations often happen at locations with a high level of information (Ewart and Johnson, 2021; Feng et al., 2022a; Orellana and Al Sayed, 2013), we find the number of hesitations increases with the number of floor signs that participants pass. Fourth, we find that straight stretches along a route has a negative impact on hesitation. It means that participants who are more persistent about the travel direction require less observation and result in fewer hesitations.

Regarding personal characteristics, we find females make more hesitations than males, which is identified by other studies (Munion et al., 2019; Suzer et al., 2018). Moreover, we find more experience with VR and computer gaming cause fewer hesitations, which is possibly due to the positive impact of familiarity with the computer system. Furthermore, the model result suggests that familiarity with the environment can reduce the number of hesitations. A similar finding is confirmed by Ewart and Johnson (2021). This is indicated by the negative impact of education_MSc, which represents the participants who are PhD students in the building.

Table 8. Three MLR models featuring the number of hesitations.

| | **MLR infra** | | | **MLR personal char** | | | **MLR infra + personal char*** | | |
|---|---|---|---|---|---|---|---|---|---|
| | **Param** | **Std** | **p-value** | **Beta** | **Std** | **p-value** | **Beta** | **Std** | **p-value** |
| Constant | -34.60 | 12.43 | <0.01 | 239.210 | 54.35 | <0.01 | 152.76 | 52.72 | <0.01 |
| Gender (male =1) | | | | -27.27 | 7.15 | <0.01 | -21.44 | 6.80 | <0.01 |
| Education_MSc | | | | 32.00 | 11.09 | <0.01 | 32.41 | 10.40 | <0.01 |
| Gaming_often | | | | -16.97 | 9.24 | 0.07 | -17.35 | 8.61 | 0.05 |
| VR_sometimes | | | | -17.17 | 5.76 | <0.01 | -17.36 | 5.41 | <0.01 |
| Height | | | | -0.99 | 0.32 | <0.01 | -0.9 | 0.30 | <0.01 |
| Task_2 | -82.37 | 28.76 | <0.01 | | | | -101.48 | 22.84 | <0.01 |
| Task_3 | -95.69 | 20.07 | <0.01 | | | | -66.35 | 16.95 | <0.01 |
| Dist_avg_straight | -1.60 | 0.00 | <0.01 | | | | -1.40 | 0.00 | <0.01 |
| Turns_left | 35.77 | 9.69 | <0.01 | | | | 20.15 | 7.72 | <0.01 |
| Turns_right | -20.89 | 9.44 | 0.03 | | | | -25.47 | 8.40 | <0.01 |
| Floorsigns | 79.93 | 14.94 | 0.06 | | | | 53.56 | 11.50 | <0.01 |
| Rot_abs | -0.10 | 0.05 | 0.05 | | | | | | |
| Firedoor | -19.12 | 9.76 | <0.01 | | | | | | |
| Adj. R square | | 0.149 | | | 0.131 | | | 0.247 | |
| F stat | | 7.088 | | | 9.421 | | | 9.29 | |
| Significance | | <0.01 | | | <0.001 | | | <0.001 | |

* Best fit model

### 5.3.2 Results and discussion of the head rotation model

Table 9 presents the results of the head rotation model. The explanatory power of the head rotation model is much better than the hesitation model. Approximately 80% of the variance is explained by the combined model that features both infrastructure and personal characteristics. Interestingly, all three models have similar explanatory power. This suggests that some of the route characteristics, such as ratio_wide, rot_abs, and percentage of windows, act as a proxy for some of the personal characteristics, and vice versa.

When studying the best-fit model (i.e., the combined model), we see that the average head rotation changes of participants decrease with the number of fire doors along the route, age, and type of task (tasks 2, 3). The results suggest older participants made fewer head rotations during the wayfinding process compared to younger participants. The studies of Kirasic (2000) and Lee and Kline (2011) confirmed the same finding, indicating that older adults require less and distinct environmental information compared to younger adults. The negative influence of the tasks might be explained by the learning effect, i.e., in the later assignments there is less of a requirement to acquire new information about the building, leading participants to rotate their heads less.

Head rotation increases with the number of floor signs along the route and self-reported orientation ability. Potentially, the first is the result of people's need to turn their heads while walking to read signs that are parallel to the corridor. The positive impact of orientation ability we find more difficult to explain. It can be that participants with better orientation skills are better at acquiring information in the environment by observation. Yet, more research is needed to better understand this mechanism. Besides that, participants with a BSc degree have a higher average head rotation. These participants are master students who have less experience in the building compared to other participants (e.g., staff members and PhD students).

Table 9. Three MLR models featuring the average head rotation.

| | MLR infra | | | MLR personal char | | | MLR infra + personal char* | | |
|---|---|---|---|---|---|---|---|---|---|
| | **Param** | **Std** | **p-value** | **Beta** | **Std** | **p-value** | **Beta** | **Std** | **p-value** |
| Constant | 41.82 | 1.48 | <0.01 | 11.56 | 2.59 | <0.01 | 31.05 | 1.79 | <0.01 |
| Ratio_wide | -24.36 | 2.94 | <0.01 | | | | | | |
| Rot_abs | 0.01 | 0.00 | <0.01 | | | | | | |
| Window | -12.65 | 7.38 | 0.09 | | | | | | |
| Firedoor | 3.10 | 0.41 | <0.01 | | | | -2.47 | 0.88 | <0.01 |
| Floorsigns | | | | | | | 2.67 | 0.00 | <0.01 |
| Age | | | | -0.17 | 0.07 | 0.01 | -0.11 | 1.29 | 0.06 |
| Gender | | | | -1.95 | 0.95 | 0.04 | | | |
| Education_BSc | | | | 1.89 | 0.61 | <0.01 | 1.49 | 0.60 | 0.01 |
| VR_often | | | | -1.27 | 0.75 | 0.10 | | | |
| Orientation_good | | | | 1.72 | 0.66 | 0.01 | 1.81 | -0.63 | 0.01 |
| Task_2 | | | | 3.96 | 0.84 | <0.01 | -11.86 | 2.47 | <0.01 |
| Task_3 | | | | 6.85 | 0.84 | <0.01 | -13.56 | 1.73 | <0.01 |
| Task_4 | | | | 24.3 | 0.84 | <0.01 | | | |
| Adj. R square | | 0.78 | | | 0.79 | | | 0.80 | |
| F stat | | 244.89 | | | 126.59 | | | 131.17 | |
| Significance | | <0.001 | | | <0.001 | | | <0.001 | |

* Best fit model

## 6. CONCLUSION AND FUTURE WORK

This study modeled pedestrian wayfinding behavior in a multi-story building at three distinct levels, starring route choice, wayfinding performance, and observation behavior. Data from a VR experiment featuring wayfinding in a multi-story building were used to develop four distinct sets of models, including one set of Path-size and Multinomial Logit models featuring the route choice behavior and three sets of Multivariate Linear regression models featuring wayfinding performance (i.e., average walking speed), hesitation behavior, and gazing behavior (i.e., head rotation). Appendix 4 provides a summary of the impact of the personal, route, and infrastructure characteristics on pedestrian wayfinding behavior regarding route choice, wayfinding performance, and observation behavior. This comprehensive modeling endeavor shows that the ***route choice*** behavior is quite consistent across participants. That is, irrespective of age, gender, orientation abilities, and gaming experience, participants prefer similar types of routes in buildings. The chosen routes tend to minimize the total distance while featuring long straight stretches and overlap with many possible routes from A to B. Whereas the ***wayfinding performance*** is predominately influenced by personal characteristics and route characteristics. Participants who are male, had more experience with computer gaming, and were more familiar with the building had better wayfinding performance. Moreover, opposite to the positive impact of straight paths on wayfinding performance, distractions such as landmarks or guidance information along the route can negatively impact wayfinding performance. ***Observation behavior*** is predominantly governed by the complexity of the task, the characteristics of the participant, and local properties of the infrastructure that provide route information (e.g., floor signs). Meanwhile, computer gaming and VR experience are among the significant variables, indicating that familiarity with computer systems should be taken into account when designing VR experiments to measure observation behavior during wayfinding.

There are also some limitations to this study. First, this paper derived observation data from head rotation data. In future studies, more precise eye gazing data (e.g., eye rotation and eye fixation duration) should be collected (e.g., via eye-tracking devices) to further substantiate the findings of the head rotation model. Second, this is the first time that we model four metrics to describe wayfinding behavior in such a comprehensive manner. During this modeling endeavor, we found some results that we cannot directly explain, such as the impact of self-reported orientation ability and education level on wayfinding performance and head rotation. More in-depth research is required to further study these effects. Third, conducting similar wayfinding experiments in other types of building networks is needed. This will provide a comparison regarding the extent to which the shape and connectivity of the building network influence the modeling results.

## REFERENCE

Afrooz, A., White, D., Parolin, B., 2018. Effects of active and passive exploration of the built environment on memory during wayfinding. Appl. Geogr. 101, 68–74. https://doi.org/10.1016/j.apgeog.2018.10.009

Agrawal, A.W., Schlossberg, M., Irvin, K., 2008. How far, by which route and why? A spatial analysis of pedestrian preference. J. Urban Des. 13, 81–98. https://doi.org/10.1080/13574800701804074

Aksoy, E., Aydin, D., İskifoğlu, G., 2020. Analysis of the Correlation Between Layout and Wayfinding Decisions in Hospitals. MEGARON 15, 509–520. https://doi.org/10.14744/megaron.2020.21797

Andree, K., Nilsson, D., Eriksson, J., 2015. Evacuation experiments in a virtual reality high-rise building: exit a and waiting time for evacuation elevators. FIRE Mater. 4B. https://doi.org/10.1002/fam

Bae, Y.H., Kim, Y.C., Oh, R.S., Son, J.Y., Hong, W.H., Choi, J.H., 2021. Walking speed reduction rates at intersections while wayfinding indoors: An experimental study. Fire Mater. 45, 498–507. https://doi.org/10.1002/fam.2821

Bae, Y.H., Kim, Y.C., Oh, R.S., Son, J.Y., Hong, W.H., Choi, J.H., 2020. Gaze point in the evacuation drills: Analysis of eye movement at the indoor wayfinding. Sustain. 12, 1–14. https://doi.org/10.3390/su12072902

Bafatakis, C., Duives, D., Daamen, W., 2015. Determining a Pedestrian Route Choice Model Through a Photo Survey. Transp. Res. Board 94th Annu. Meet.

Basu, R., Sevtsuk, A., Li, X., 2022. How do street attributes affect willingness-to-walk? City-wide pedestrian route choice analysis using big data from Boston and San Francisco. Transp. Res. A in review, 1–19. https://doi.org/10.1016/j.tra.2022.06.007

Benthorn, L., Frantzich, H., 1999. Fire alarm in a public building: How do people evaluate information and choose an evacuation exit? Fire Mater. 23, 311–315. https://doi.org/10.1002/(SICI)1099-1018(199911/12)23:6<311::AID-FAM704>3.0.CO;2-J

Blue, V.J., Adler, J.L., 2001. Cellular automata microsimulation for modeling bi-directional pedestrian walkways. Transp. Res. Part B Methodol. 35, 293–312. https://doi.org/10.1016/S0191-2615(99)00052-1

Borgers, A., Timmermans, H., 1986. A model of pedestrian route choice and demand for retail facilities within inner-city shopping areas. Geogr. Anal. 18, 115–128.

Borst, H.C., de Vries, S.I., Graham, J.M.A., van Dongen, J.E.F., Bakker, I., Miedema, H.M.E., 2009. Influence of environmental street characteristics on walking route choice of elderly people. J. Environ. Psychol. 29, 477–484. https://doi.org/10.1016/j.jenvp.2009.08.002

Bosina, E., Weidmann, U., 2017. Estimating pedestrian speed using aggregated literature data. Phys. A Stat. Mech. its Appl. 468, 1–29. https://doi.org/10.1016/j.physa.2016.09.044

Broach, J., Dill, J., 2015. Pedestrian Route Choice Model Estimated from Revealed Preference GPS Data. Transp. Res. Board 93rd Annu. Meet. January 12-16, Washington, D.C.

Brooke, J., 1996. SUS - A quick and dirty usability scale. Usability Eval. Ind. 189, 4–7. https://doi.org/10.1002/hbm.20701

Brown, B.B., Werner, C.M., Amburgey, J.W., Szalay, C., 2007. Walkable route perceptions and physical features: Converging evidence for en route walking experiences. Environ. Behav. 39, 34–61. https://doi.org/10.1177/0013916506295569

Cai, L., Yang, R., Tao, Z., 2018. A new method of evaluating signage system using mixed reality and eye tracking, in: Proceedings of the 4th ACM SIGSPATIAL International Workshop on Safety and Resilience, EM-GIS 2018. https://doi.org/10.1145/3284103.3284105

Cao, L., Lin, J., Li, N., 2019. A virtual reality based study of indoor fire evacuation after active or passive spatial exploration. Comput. Human Behav. 90, 37–45. https://doi.org/10.1016/j.chb.2018.08.041

Cao, S., Fu, L., Wang, P., Zeng, G., Song, W., 2018. Experimental and modeling study on evacuation under good and limited visibility in a supermarket. Fire Saf. J. 102, 27–36. https://doi.org/10.1016/j.firesaf.2018.10.003

Castelli, L., Latini Corazzini, L., Geminiani, G.C., 2008. Spatial navigation in large-scale virtual environments: Gender differences in survey tasks. Comput. Human Behav. 24, 1643–1667. https://doi.org/10.1016/j.chb.2007.06.005

Chen, C.H., Chang, W.C., Chang, W. Te, 2009. Gender differences in relation to wayfinding strategies, navigational support design, and wayfinding task difficulty. J. Environ. Psychol. 29, 220–226. https://doi.org/10.1016/j.jenvp.2008.07.003

Chen, J., Li, N., Shi, Y., Du, J., 2023. Cross-cultural assessment of the effect of spatial information on firefighters' wayfinding performance: A virtual reality-based study. Int. J. Disaster Risk Reduct. 84, 103486. https://doi.org/10.1016/j.ijdrr.2022.103486

Chen, L., Tang, T.Q., Huang, H.J., Song, Z., 2018. Elementary students' evacuation route choice in a classroom: A questionnaire-based method. Phys. A Stat. Mech. its Appl. 492, 1066–1074. https://doi.org/10.1016/j.physa.2017.11.036

Cheung, C.Y., Lam, W.H.K., 1998. Pedestrian route choices between escalator and stairway in MTR stations. J. Transp. Eng. 124, 277–285. https://doi.org/10.1061/(ASCE)0733-947X(1998)124:3(277)

Choi, J., Galea, E.R., Hong, W., 2014. Individual Stair Ascent and Descent Walk Speeds Measured in a Korean High-Rise Building. Fire Technol. 50, 267–295. https://doi.org/10.1007/s10694-013-0371-4

Cliburn, D.C., Rilea, S.L., 2008. Showing users the way: Signs in virtual worlds. Proc. - IEEE Virtual Real. 129–132. https://doi.org/10.1109/VR.2008.4480763

Conroy, R., 2001. Spatial Navigation in Immersive Virtual Environments. University of London.

Dalton, R.C., 2003. The secret is to follow your nose: Route path selection and angularity. Environ. Behav. 35, 107–131. https://doi.org/10.1177/0013916502238867

Dias, C., Ejtemai, O., Sarvi, M., Shiwakoti, N., 2014. Pedestrian Walking Characteristics Through Angled Corridors. Transp. Res. Rec. J. Transp. Res. Board 2421, 41–50. https://doi.org/10.3141/2421-05

Dietrich, F., Köster, G., 2014. Gradient navigation model for pedestrian dynamics. Phys. Rev. E - Stat. Nonlinear, Soft Matter

Phys. 89, 1–19. https://doi.org/10.1103/PhysRevE.89.062801
Dijkstra, E.W., 1959. A Note on Two Problems in Connexion with Graphs, in: Edsger Wybe Dijkstra: His Life, Work, and Legacy. pp. 287–290. https://doi.org/10.1145/3544585.3544600
Dogu, U., Erkip, F., 2000. Spatial factors affecting wayfinding and orientation: A case study in a shopping mall. Environ. Behav. 32, 731–755. https://doi.org/10.1177/00139160021972775
Dong, W., Qin, T., Liao, H., Liu, Y., Liu, J., 2020. Comparing the roles of landmark visual salience and semantic salience in visual guidance during indoor wayfinding. Cartogr. Geogr. Inf. Sci. 47, 229–243. https://doi.org/10.1080/15230406.2019.1697965
Duarte, E., Rebelo, F., Teles, J., Wogalter, M.S., 2014. Behavioral compliance for dynamic versus static signs in an immersive virtual environment. Appl. Ergon. 45, 1367–1375. https://doi.org/10.1016/j.apergo.2013.10.004
Duives, D., Mahmassani, H., 2012. Exit Choice Decisions During Pedestrian Evacuations of Buildings. Transp. Res. Rec. J. Transp. Res. Board 2316, 84–94. https://doi.org/10.3141/2316-10
Duncan, M.J., Mummery, W.K., 2007. GIS or GPS? A Comparison of Two Methods For Assessing Route Taken During Active Transport. Am. J. Prev. Med. 33, 51–53. https://doi.org/10.1016/j.amepre.2007.02.042
Ewart, I.J., Johnson, H., 2021. Virtual reality as a tool to investigate and predict occupant behaviour in the real world: The example of wayfinding. J. Inf. Technol. Constr. 26, 286–302. https://doi.org/10.36680/j.itcon.2021.016
Fang, Z., Song, W., Zhang, J., Wu, H., 2010. Experiment and modeling of exit-selecting behaviors during a building evacuation. Physica A 389, 815–824. https://doi.org/10.1016/j.physa.2009.10.019
Feng, Y., Duives, D., Daamen, W., Hoogendoorn, S., 2021a. Data collection methods for studying pedestrian behaviour: A systematic review. Build. Environ. 187, 107329. https://doi.org/https://doi.org/10.1016/j.buildenv.2020.107329
Feng, Y., Duives, D., Hoogendoorn, S., 2022a. Development and evaluation of a VR research tool to study wayfinding behaviour in a multi-story building. Saf. Sci. 147, 105573. https://doi.org/10.1016/j.ssci.2021.105573
Feng, Y., Duives, D.C., Hoogendoorn, S.P., 2022b. Wayfinding behaviour in a multi-level building : a comparative study of HMD VR and Desktop VR. Adv. Eng. Informatics 51, 101475. https://doi.org/10.1016/j.aei.2021.101475
Feng, Y., Duives, D.C., Hoogendoorn, S.P., 2021b. Using virtual reality to study pedestrian exit choice behaviour during evacuations. Saf. Sci. 137, 105158. https://doi.org/10.1016/j.ssci.2021.105158
Ferrer, S., Ruiz, T., Mars, L., 2015. A qualitative study on the role of the built environment for short walking trips. Transp. Res. Part F Psychol. Behav. 33, 141–160. https://doi.org/10.1016/j.trf.2015.07.014
Fitzpatrick, K., Brewer, M.A., Turner, S., 2006. Another Look at Pedestrian Walking Speed. Transp. Res. Rec. 1982, 21–29.
Fossum, M., Ryeng, E.O., 2022. Pedestrians' and bicyclists' route choice during winter conditions. Urban, Plan. Transp. Res. 10, 38–57. https://doi.org/10.1080/21650020.2022.2034524
Fridolf, K., Ronchi, E., Nilsson, D., Frantzich, H., 2013. Movement speed and exit choice in smoke- filled rail tunnels. Fire Saf. J. 59, 8–21. https://doi.org/10.1016/j.firesaf.2013.03.007
Galea, E.R., Xie, H., Deere, S., Cooney, D., Filippidis, L., 2017. An international survey and full-scale evacuation trial demonstrating the effectiveness of the active dynamic signage system concept. Fire Mater. 41, 493–513. https://doi.org/10.1002/fam.2414
Gärling, T., Lindberg, E., Mäntylä, T., 1983. Orientation in buildings: Effects of familiarity, visual access, and orientation aids. J. Appl. Psychol. 68, 177–186. https://doi.org/10.1037/0021-9010.68.1.177
Goldiez, B.F., Ahmad, A.M., Hancock, P.A., 2007. Effects of augmented reality display settings on human wayfinding performance. IEEE Trans. Syst. Man Cybern. Part C Appl. Rev. 37, 839–845. https://doi.org/10.1109/TSMCC.2007.900665
Guo, Z., 2009. Does the pedestrian environment affect the utility of walking? A case of path choice in downtown Boston. Transp. Res. Part D Transp. Environ. 14, 343–352. https://doi.org/10.1016/j.trd.2009.03.007
Guo, Z., Ferreira, J., 2008. Pedestrian environments, transit path choice, and transfer penalties: Understanding land-use impacts on transit travel. Environ. Plan. B Plan. Des. 35, 461–479. https://doi.org/10.1068/b33074
Haghani, M., Sarvi, M., 2016. Pedestrian crowd tactical-level decision making during emergency evacuations. J. Adv. Transp. 50, 1870–1895. https://doi.org/10.1002/atr.1434
Hamouni, P., 2018. A Pedestrian Route Choice Model Concerning Quantified Built Environment Factors. Concordia University.
Heino, T., Cliburn, D., Rilea, S., Cooper, J., Tachkov, V., 2010. Limitations of signs as navigation aids in virtual worlds. Proc. Annu. Hawaii Int. Conf. Syst. Sci. 1–10. https://doi.org/10.1109/HICSS.2010.256
Helbing, D., Molnár, P., 1995. Social force model for pedestrian dynamics. Phys. Rev. E 51, 4282–4286. https://doi.org/10.1103/PhysRevE.51.4282
Heliövaara, S., Kuusinen, J.M., Rinne, T., Korhonen, T., Ehtamo, H., 2012. Pedestrian behavior and exit selection in evacuation of a corridor - An experimental study. Saf. Sci. 50, 221–227. https://doi.org/10.1016/j.ssci.2011.08.020
Hintaran, R.E., 2015. Unravelling Urban Pedestrian Trips: Developing a new pedestrian route choice model estimated from revealed preference GPS data.
Hölscher, C., Meilinger, T., Vrachliotis, G., Brösamle, M., Knauff, M., 2007. Up the down staircase : Wayfinding strategies in multi-level buildings. J. Environ. Psychol. 26, 284–299. https://doi.org/10.1016/j.jenvp.2006.09.002
Hood, J., Sall, E., Charlton, B., 2011. A GPS-based bicycle route choice model for San Francisco, California. Transp. Lett. 3, 63–75. https://doi.org/10.3328/TL.2011.03.01.63-75
Hoogendoorn-Lanser, S., 2005. Modelling Travel Behaviour in Multi-modal Networks. TRAIL Thesis Ser.
Hoogendoorn, S.P., Van Wageningen-Kessels, F.L.M., Daamen, W., Duives, D.C., 2014. Continuum modelling of pedestrian

flows: From microscopic principles to self-organised macroscopic phenomena. Phys. A Stat. Mech. its Appl. 416, 684–694. https://doi.org/10.1016/j.physa.2014.07.050
Iachini, T., Ruotolo, F., Ruggiero, G., 2009. The effects of familiarity and gender on spatial representation. J. Environ. Psychol. 29, 227–234. https://doi.org/10.1016/j.jenvp.2008.07.001
Isenschmid, U.;, Widmer, A.;, Meister, A.;, Felder, M.;, Axhausen, K.W., 2022. A Zurich pedestrian route choice model based on BFSLE choice set generation Working Paper.
Jansen-Osmann, P., Wiedenbauer, G., 2004. Wayfinding performance in and the spatial knowledge of a color-coded building for adults and children. Spat. Cogn. Comput. 4, 337–358. https://doi.org/10.1207/s15427633scc0404_3
Jin, L., Lu, W., Sun, P., 2022. Effect of the Street Environment on Walking Behavior: A Case Study Using the Route Choice Model in the Chunliu Community of Dalian. Front. Public Heal. 10, 1–11. https://doi.org/10.3389/fpubh.2022.874788
Kalantari, S., Tripathi, V., Kan, J., Rounds, J.D., Mostafavi, A., Snell, R., Cruz-Garza, J.G., 2022. Evaluating the impacts of color, graphics, and architectural features on wayfinding in healthcare settings using EEG data and virtual response testing. J. Environ. Psychol. 79, 101744. https://doi.org/10.1016/j.jenvp.2021.101744
Kaptein, N.A., Theeuwes, J., van der Horst, R., 1996. Driving simulator validity: some considerations. Transp. Res. Rec. 30–36. https://doi.org/10.3141/1550-05
Kato, Y., Takeuchi, Y., 2003. Individual differences in wayfinding strategies. J. Environ. Psychol. 23, 171–188. https://doi.org/10.1016/S0272-4944(03)00011-2
Kennedy, R.S., Lane, N.E., Berbaum, K.S., Lilienthal, M.G., 1993. Simulator Sickness Questionnaire: An Enhanced Method for Quantifying Simulator Sickness. Int. J. Aviat. Psychol. 3, 203–220. https://doi.org/10.1207/s15327108ijap0303_3
Kinateder, M., Comunale, B., Warren, W.H., 2018. Exit choice in an emergency evacuation scenario is influenced by exit familiarity and neighbor behavior. Saf. Sci. 106, 170–175. https://doi.org/10.1016/j.ssci.2018.03.015
Kirasic, K.C., 2000. Age differences in adults' spatial abilities, learning environmental layout, and wayfinding behavior. Spat. Cogn. Comput. 2, 117–134.
Kobes, M., Helsloot, I., De Vries, B., Post, J., 2010a. Exit choice, (pre-)movement time and (pre-)evacuation behaviour in hotel fire evacuation - Behavioural analysis and validation of the use of serious gaming in experimental research. Procedia Eng. 3, 37–51. https://doi.org/10.1016/j.proeng.2010.07.006
Kobes, M., Helsloot, I., de Vries, B., Post, J.G., Oberijé, N., Groenewegen, K., 2010b. Way finding during fire evacuation; an analysis of unannounced fire drills in a hotel at night. Build. Environ. 45, 537–548. https://doi.org/10.1016/j.buildenv.2009.07.004
Koh, P.P., Wong, Y.D., 2013. Influence of infrastructural compatibility factors on walking andcycling route choices. J. Environ. Psychol. 36, 202–213. https://doi.org/10.1016/j.jenvp.2013.08.001
Kuliga, S.F., Nelligan, B., Dalton, R.C., Marchette, S., Shelton, A.L., Carlson, L., Hölscher, C., 2019. Exploring Individual Differences and Building Complexity in Wayfinding: The Case of the Seattle Central Library. Environ. Behav. 51, 622–665. https://doi.org/10.1177/0013916519836149
Lee, S., Kline, R., 2011. Wayfinding study in virtual environments: The elderly vs. the younger-aged groups. ArchNet-IJAR Int. J. Archit. Res. 5, 63.
Li, H., Thrash, T., Hölscher, C., Schinazi, V.R., 2019. The effect of crowdedness on human wayfinding and locomotion in a multi- level virtual shopping mall. J. Environ. Psychol. 65, 101320. https://doi.org/10.1016/j.jenvp.2019.101320
Lin, J., Cao, L., Li, N., 2019. Assessing the influence of repeated exposures and mental stress on human wayfinding performance in indoor environments using virtual reality technology. Adv. Eng. Informatics 39, 53–61. https://doi.org/10.1016/j.aei.2018.11.007
Lin, J., Zhu, R., Li, N., Becerik-Gerber, B., 2020. Do people follow the crowd in building emergency evacuation? A cross-cultural immersive virtual reality-based study. Adv. Eng. Informatics 43, 101040. https://doi.org/10.1016/j.aei.2020.101040
Lin, P., Gao, D. li, Wang, G.Y., Wu, F.Y., Ma, J., Si, Y.L., Ran, T., 2019. The Impact of an Obstacle on Competitive Evacuation Through a Bottleneck. Fire Technol. 55, 1967–1981. https://doi.org/10.1007/s10694-019-00838-4
Liu, Y., Yang, D., Timmermans, H.J.P., de Vries, B., 2020. The impact of the street-scale built environment on pedestrian metro station access/egress route choice. Transp. Res. Part D Transp. Environ. 87, 102491. https://doi.org/10.1016/j.trd.2020.102491
López-Lambas, M.E., Sánchez, J.M., Alonso, A., 2021. The walking health: A route choice model to analyze the street factors enhancing active mobility. J. Transp. Heal. 22. https://doi.org/10.1016/j.jth.2021.101133
Lovreglio, R., Dillies, E., Kuligowski, E., Rahouti, A., Haghani, M., 2021. Investigating Exit Choice in Built Environment Evacuation combining Immersive Virtual Reality and Discrete Choice Modelling. https://doi.org/10.1016/j.autcon.2022.104452
Lue, G., Miller, E.J., 2019. Estimating a Toronto pedestrian route choice model using smartphone GPS data. Travel Behav. Soc. 14, 34–42. https://doi.org/10.1016/j.tbs.2018.09.008
Mackay Yarnal, C.M., Coulson, M.R.C., 1982. Recreational Map Design and Map Use: An Experiment. Cartogr. J. 19, 16–27. https://doi.org/10.1179/caj.1982.19.1.16
Malinowski, J.C., Gillespie, W.T., 2001. Individual differences in performance on a large-scale, real-world wayfinding task. J. Environ. Psychol. 21, 73–82. https://doi.org/10.1006/jevp.2000.0183
Meng, F., Zhang, W., 2014. Way-finding during a fire emergency: An experimental study in a virtual environment. Ergonomics. https://doi.org/10.1080/00140139.2014.904006

Moeser, S.D., 1988. Cognitive mapping in a complex building. Environ. Behav. https://doi.org/10.1177/0013916588201002
Moffat, S.D., Zonderman, A.B., Resnick, S.M., 2001. Age differences in spatial memory in a virtual environment navigation task. Neurobiol. Aging. https://doi.org/10.1016/S0197-4580(01)00251-2
Montello, D.R., Sas, C., 2006. Human Factors of Wayfinding in Navigation. Hum. Factors 2003–2008.
Montini, L., Antoniou, C., Axhausen, K.W., 2017. Route and mode choice models using GPS data. TRB Annu. Meet. Online 1204, 17–03082.
Montini, L., Axhausen, K.W., 2015. Preliminary results: Route choice analysis from multi-day GPS data. 15th Swiss Transp. Res. Conf.
Morganti, F., Carassa, A., Geminiani, G., 2007. Planning optimal paths: A simple assessment of survey spatial knowledge in virtual environments. Comput. Human Behav. 23, 1982–1996. https://doi.org/10.1016/j.chb.2006.02.006
Munion, A.K., Stefanucci, J.K., Rovira, E., Squire, P., Hendricks, M., 2019. Gender differences in spatial navigation: Characterizing wayfinding behaviors. Psychon. Bull. Rev. 26, 1933–1940. https://doi.org/10.3758/s13423-019-01659-w
Muraleetharan, T., Hagiwara, T., 2007. Overall Level of Service of Urban Walking Environment and Its Influence on Pedestrian Route Choice Behavior. Transp. Res. Rec. J. Transp. Res. Board 2002, 7–17. https://doi.org/10.3141/2002-02
Murias, K., Kwok, K., Castillejo, A.G., Liu, I., Iaria, G., 2016. The effects of video game use on performance in a virtual navigation task. Comput. Human Behav. 58, 398–406. https://doi.org/10.1016/j.chb.2016.01.020
Nam, C.S., Whang, M., Liu, S., Moore, M., 2015. Wayfinding of Users With Visual Impairments in Haptically Enhanced Virtual Environments. Int. J. Hum. Comput. Interact. 31, 295–306. https://doi.org/10.1080/10447318.2015.1004151
Natapov, A., Fisher-Gewirtzman, D., 2016. Visibility of urban activities and pedestrian routes: An experiment in a virtual environment. Comput. Environ. Urban Syst. 58, 60–70. https://doi.org/10.1016/j.compenvurbsys.2016.03.007
Nori, R., Piccardi, L., Maialetti, A., Goro, M., Rossetti, A., Argento, O., Guariglia, C., 2018. No gender differences in egocentric and allocentric environmental transformation after compensating for male advantage by manipulating familiarity. Front. Neurosci. 12, 1–9. https://doi.org/10.3389/fnins.2018.00204
O'Neill, M.J., 1992. Effects of familiarity and plan complexity on wayfinding in simulated buildings. J. Environ. Psychol. 12, 319–327.
O'Neill, M.J., 1991. Effects of signage and floor plan configuration on wayfinding accuracy. Environ. Behav. 23, 553–574. https://doi.org/10.1177/0013916591235002
Ohm, C., Müller, M., Ludwig, B., 2017. Evaluating indoor pedestrian navigation interfaces using mobile eye tracking. Spat. Cogn. Comput. 17, 89–120. https://doi.org/10.1080/13875868.2016.1219913
Omer, I., Goldblatt, R., 2007. The implications of inter-visibility between landmarks on wayfinding performance: An investigation using a virtual urban environment. Comput. Environ. Urban Syst. 31, 520–534. https://doi.org/10.1016/j.compenvurbsys.2007.08.004
Orellana, N., Al Sayed, K., 2013. On spatial wayfinding: Agent and human navigation patterns in virtual and real worlds. 2013 Int. Sp. Syntax Symp.
Peponis, J., Zimring, C., Choi, Y.K., 1990. Finding the building in wayfinding.pdf. Environ. Behav. 22, 555–590.
Pouyan, A.E., Ghanbaran, A., Shakibamanesh, A., 2021. Impact of circulation complexity on hospital wayfinding behavior (Case study: Milad 1000-bed hospital, Tehran, Iran). J. Build. Eng. 44, 102931. https://doi.org/10.1016/j.jobe.2021.102931
Rieser-Schüssler, N., Balmer, M., Axhausen, K.W., 2013. Route choice sets for very high-resolution data. Transp. A Transp. Sci. 9, 825–845. https://doi.org/10.1080/18128602.2012.671383
Rodríguez, D.A., Merlin, L., Prato, C.G., Conway, T.L., Cohen, D., Elder, J.P., Evenson, K.R., McKenzie, T.L., Pickrel, J.L., Veblen-Mortenson, S., 2015. Influence of the Built Environment on Pedestrian Route Choices of Adolescent Girls, Environment and Behavior. https://doi.org/10.1177/0013916513520004
Ruddle, R.A., Lessels, S., 2006. Three levels of metric for evaluating wayfinding, in: Presence: Teleoperators and Virtual Environments. pp. 637–654. https://doi.org/10.1162/pres.15.6.637
Schrom-Feiertag, H., Settgast, V., Seer, S., 2017. Evaluation of indoor guidance systems using eye tracking in an immersive virtual environment. Spat. Cogn. Comput. 17, 163–183. https://doi.org/10.1080/13875868.2016.1228654
Seneviratne, P.N., Morrall, J.F., 1985. Analysis of factors affecting the choice of route of pedestrians. Transp. Plan. Technol. 10, 147–159. https://doi.org/10.1080/03081068508717309
Sevtsuk, A., Basu, R., 2022. The role of turns in pedestrian route choice: A clarification. J. Transp. Geogr. 102, 103392. https://doi.org/10.1016/j.jtrangeo.2022.103392
Sevtsuk, A., Basu, R., Li, X., Kalvo, R., 2021. A big data approach to understanding pedestrian route choice preferences: Evidence from San Francisco. Travel Behav. Soc. 25, 41–51. https://doi.org/10.1016/j.tbs.2021.05.010
Shi, Y., Kang, J., Xia, P., Tyagi, O., Mehta, R.K., Du, J., 2021a. Spatial knowledge and firefighters' wayfinding performance: A virtual reality search and rescue experiment. Saf. Sci. 105231. https://doi.org/10.1016/j.ssci.2021.105231
Shi, Y., Kang, J., Xia, P., Tyagi, O., Mehta, R.K., Du, J., 2021b. The Role of Spatial Information in Search and Rescue: A Virtual Reality Experiment, in: Computing in Civil Engineering2021. pp. 1285–1292. https://doi.org/10.1061/9780784483893.157
Shields, T.J., Boyce, K.E., 2000. A study of evacuation from large retail stores. Fire Saf. J. 35, 25–49. https://doi.org/10.1016/S0379-7112(00)00013-8
Shiwakoti, N., Tay, R., Stasinopoulos, P., Woolley, P.J., 2017. Likely behaviours of passengers under emergency evacuation in train station. Saf. Sci. 91, 40–48. https://doi.org/10.1016/j.ssci.2016.07.017
Silva, J.F., Almeida, J.E., Rossetti, R.J.F., Coelho, A.L., 2013. A serious game for EVAcuation training, in: SeGAH 2013 - IEEE 2nd International Conference on Serious Games and Applications for Health, Book of Proceedings. pp. 1–6.

https://doi.org/10.1109/SeGAH.2013.6665302
Slone, E., Burles, F., Robinson, K., Levy, R.M., Iaria, G., 2015. Floor Plan Connectivity Influences Wayfinding Performance in Virtual Environments. Environ. Behav. 47, 1024–1053. https://doi.org/10.1177/0013916514533189
Smith, S.P., Du'Mont, S., 2009. Measuring the effect of gaming experience on virtual environment navigation tasks. 3DUI - IEEE Symp. 3D User Interfaces 2009 - Proc. 3–10. https://doi.org/10.1109/3DUI.2009.4811198
Soeda, M., Kushiyama, N., Ryuzo, O., 1997. Wayfinding in cases with vertical motion, in: Proceedings of MERA 97. pp. 559–564.
Soh, B.K., Smith-Jackson, T.L., 2004. Influence of map design, individual differences, and environmental cues on wayfinding performance. Spat. Cogn. Comput. 4, 137–165. https://doi.org/10.1207/s15427633scc0402_2
Stubenschrott, M., Kogler, C., Matyus, T., Seer, S., 2014. A dynamic pedestrian route choice model validated in a high density subway station. Transp. Res. Procedia 2, 376–384. https://doi.org/10.1016/j.trpro.2014.09.036
Suzer, O.K., Olgunturk, N., Guvenc, D., 2018. The effects of correlated colour temperature on wayfinding: A study in a virtual airport environment. Displays 51, 9–19. https://doi.org/10.1016/j.displa.2018.01.003
Taillade, M., Sauzéon, H., Dejos, M., Arvind Pala, P., Larrue, F., Wallet, G., Gross, C., N'Kaoua, B., 2013. Executive and memory correlates of age-related differences in wayfinding performances using a virtual reality application. Aging, Neuropsychol. Cogn. 20, 298–319. https://doi.org/10.1080/13825585.2012.706247
Tascón, L., Di Cicco, C., Piccardi, L., Palmiero, M., Bocchi, A., Cimadevilla, J.M., 2021. Sex differences in spatial memory: Comparison of three tasks using the same virtual context. Brain Sci. 11. https://doi.org/10.3390/brainsci11060757
Tian, P., Wang, Yunjia, Lu, Y., Zhang, Y., Wang, X., Wang, Yong, 2019. Behavior Analysis of Indoor Escape Route-finding Based on Head-mounted VR and Eye Tracking, in: 2019 International Conference on Internet of Things (iThings) and IEEE Green Computing and Communications (GreenCom) and IEEE Cyber, Physical and Social Computing (CPSCom) and IEEE Smart Data (SmartData). IEEE, pp. 422–427. https://doi.org/10.1109/iThings/GreenCom/CPSCom/SmartData.2019.00090
Ton, D., Cats, O., Duives, D., Hoogendoorn, S., 2017. How do people cycle in amsterdam, Netherlands?: Estimating cyclists' route choice determinants with GPS data from an Urban area. Transp. Res. Rec. 2662, 75–82. https://doi.org/10.3141/2662-09
Tsukaguchi, H., Matsuda, K., 2002. Analysis on pedestrian route choice behavior. Doboku Gakkai Ronbunshu 709, 117–126.
Tsukaguchi, H., Ohashi, Y., 2007. Modelling pedestrian route choice behaviour and its application to the planning of underground shopping streets. 11th ACUUS Int. Conf. - Undergr. Sp. Expand. Front. 527–532.
Veeraswamy, A., Galea, E.R., Lawrence, P.J., 2011. Wayfinding behavior within buildings - An international survey. Fire Saf. Sci. 735–748. https://doi.org/10.3801/IAFSS.FSS.10-735
Verghote, A., Al-Haddad, S., Goodrum, P., Van Emelen, S., 2019. The effects of information format and spatial cognition on individual wayfinding performance. Buildings 9. https://doi.org/10.3390/buildings9020029
Vilar, E., Rebelo, F., Noriega, P., 2014a. Indoor Human Wayfinding Performance Using Vertical and Horizontal Signage in Virtual Reality. Hum. Factors Ergon. Manuf. Serv. Ind. 24, 601–615. https://doi.org/10.1002/hfm
Vilar, E., Rebelo, F., Noriega, P., Duarte, E., Mayhorn, C.B., 2014b. Effects of competing environmental variables and signage on route-choices in simulated everyday and emergency wayfinding situations. Ergonomics 57, 511–524. https://doi.org/10.1080/00140139.2014.895054
Vilar, E., Rebelo, F., Noriega, P., Teles, J., Mayhorn, C., 2015. Signage versus environmental affordances: Is the explicit information strong enough to guide human behavior during a wayfinding task? Hum. Factors Ergon. Manuf. 25, 439–452. https://doi.org/10.1002/hfm.20557
Wang, C., Chen, Y., Zheng, S., Liao, H., 2019. Gender and age differences in using indoor maps for wayfinding in real environments. ISPRS Int. J. Geo-Information 8. https://doi.org/10.3390/ijgi8010011
Werberich, B.R., Pretto, C.O., Cybis, H.B.B., 2015. Calibration of a pedestrian route choice model with a basis in friction forces. Transp. Res. Rec. 2519, 137–145. https://doi.org/10.3141/2519-15
Wiener, J.M., Hölscher, C., Büchner, S., Konieczny, L., 2012. Gaze behaviour during space perception and spatial decision making. Psychol. Res. 76, 713–729. https://doi.org/10.1007/s00426-011-0397-5
Witmer, B.G., Jerome, C.J., Singer, M.J., 2005. The Factor Structure of the Presence Questionnaire. Presence Teleoperators Virtual Environ. 14, 298–312.
Yang, Y., Merrill, E.C., Robinson, T., Wang, Q., 2018. The impact of moving entities on wayfinding performance. J. Environ. Psychol. 56, 20–29. https://doi.org/10.1016/j.jenvp.2018.02.003
Zhang, J., Klingsch, W., Schadschneider, A., Seyfried, A., 2011. Transitions in pedestrian fundamental diagrams of straight corridors and T-junctions. J. Stat. Mech. Theory Exp. 2011. https://doi.org/10.1088/1742-5468/2011/06/P06004
Zhang, M., Ke, J., Tong, L., Luo, X., 2021. Investigating the influence of route turning angle on compliance behaviors and evacuation performance in a virtual-reality-based experiment. Adv. Eng. Informatics 48, 101259. https://doi.org/10.1016/j.aei.2021.101259
Zhang, S., Park, S., 2021. Study of Effective Corridor Design to Improve Wayfinding in Underground Malls. Front. Psychol. 12, 1–13. https://doi.org/10.3389/fpsyg.2021.631531
Zhou, Y., Cheng, X., Zhu, L., Qin, T., Dong, W., Liu, J., 2020. How does gender affect indoor wayfinding under time pressure? Cartogr. Geogr. Inf. Sci. 47, 367–380. https://doi.org/10.1080/15230406.2020.1760940
Zhu, K.-J.J., Shi, Q., 2016. Experimental Study on Choice Behavior of Pedestrians during Building Evacuation. Procedia Eng. 135, 206–215. https://doi.org/10.1016/j.proeng.2016.01.110

Zomer, L.-B., Schneider, F., Ton, D., Hoogendoorn-Lanser, S., Duives, D., Cats, O., Hoogendoorn, S., 2019. Determinants of urban wayfinding styles. Travel Behav. Soc. 17, 72–85. https://doi.org/10.1016/j.tbs.2019.07.002

## APPENDIX

**Appendix 1. Overview of modeling studies featuring pedestrian route choices**

| Authors | Model type | Area of study | Type population | Indoor /Outdoor | No. participants | No. routes | Socio | Personal | Route | Infra | Network | Land use |
|---|---|---|---|---|---|---|---|---|---|---|---|---|
| (Fossum and Ryeng, 2022) | Binary logit | Trondheim, NO | Adults & Elderly | Outdoor | 1,677 | 1,677 | x | - | x | x | - | - |
| (Isenschmid et al., 2022) | PSL | CH | Adults | Outdoor | 202 | 922 | x | - | x | - | - | x |
| (Jin et al., 2022) | Conditional logit | Chunliu, CN | Adults | Outdoor | 219 | 219 | - | - | x | x | - | x |
| (Sevtsuk and Basu, 2022) | PSL | San Francisco/ Boston, USA | Adults | Outdoor | ? | 14,760/ 11,165 | - | - | x | x | - | x |
| (López-Lambas et al., 2021) | Statistics only | Madrid | Adults | Outdoor | 1,055 | 1,055 | - | - | x | - | - | - |
| (Sevtsuk et al., 2021) | PSL | San Francisco | Adults | Outdoor | ? | 14,760 | - | - | x | x | - | - |
| (Liu et al., 2020) | PSC /LC | Tianjin, CN | Adults | Outdoor | 402 | 402 | - | - | x | x | - | x |
| (Lue and Miller, 2019) | PSL | Toronto area, CA | Adults | Outdoor | 71 | 776 | x | x | x | x | - | x |
| (Hamouni, 2018) | PSL | Montreal, CA | Adults | Outdoor | 2414 | 1531 | - | - | x | x | - | x |
| (Montini et al., 2017) | PSL | Zurich, CH | Adults | Outdoor | 150 | 985 | - | x | x | x | - | - |
| (Broach and Dill, 2015) | MNL | US | Adults | Outdoor | ?? | ?? | - | - | x | x | x | x |
| (Bafatakis et al., 2015) | MNL | Delft, NL | Adults | Outdoor | ?? | ?? | - | - | - | x | - | - |
| (Ferrer et al., 2015) | Qualitative | Valencia, ES | Students & Adults | Outdoor | 23 | 23 | - | - | x | x | - | - |
| (Rodríguez et al., 2015) | PSL | San Diego/ Minneapolis, USA | Children | Outdoor | 33 | 112 | - | - | x | x | - | x |
| (Montini and Axhausen, 2015) | PSL | Zurich, CH | Adults | Outdoor | 150 | 985 | - | - | x | x | - | - |
| (Hintaran, 2015) | PSL | Zurich, CH | Adults | Outdoor | 51 | 580 | - | - | x | - | - | - |
| (Werberich et al., 2015) | Physics model | Experimental setting | Adults | Indoor | 40 | 40 | - | - | x | - | - | - |
| (Stubenschrott et al., 2014) | Utility max | U2 subway station, CH | Adults | Indoor | 16,500 | ? | - | - | x | - | - | - |
| (Koh and Wong, 2013) | Statistics only | Near Rail stations, Singapore | Adults | Outdoor | 1146 | 703 | - | - | x | x | - | x |
| (Borst et al., 2009) | ML regression | Schiedam, NL | Elderly | Outdoor | 364 | 364 | - | - | x | x | - | x |
| (Guo, 2009) | Binary logit | Downtown, Boston, USA | Adults | Outdoor | ? | 711 | - | x | x | x | x | - |
| (Guo and Ferreira, 2008) | Logit | Downtown, Boston, USA | Adults | Outdoor | 3140 | 3140 | - | x | x | x | x | - |
| (Agrawal et al., 2008) | Statistics only | Near rail stations, California and Oregon, USA | Adults | Outdoor | 328 | 328 | - | - | x | x | - | x |
| (Tsukaguchi and Ohashi, 2007) | Binary logit | Shopping mall, Osaka and Tokyo, JP | Adults | Indoor | 462 | 462 | - | - | x | - | - | - |
| (Duncan and Mummery, 2007) | Statistics only | Rockhampton, AU | Children | Outdoor | 75 | 75 | - | - | x | - | - | - |
| (Brown et al., 2007) | Statistics only | Downtown, Salt Lake City, USA | Students | Outdoor | 73 | 73 | x | x | x | x | - | x |
| (Muraleetharan and Hagiwara, 2007) | MNL | Sapporo, JP | Adults | Outdoor | 354 | 354 | - | - | x | x | - | - |
| (Tsukaguchi and Matsuda, 2002) | ? | ? | ? | Indoor | ? | ? | ? | ? | ? | ? | ? | ? |
| (Cheung and Lam, 1998) | Logit model | MTR stations, Hong Kong | Adults | Indoor | 38,839 | 38,839 | - | - | x | - | - | - |
| (Borgers and Timmermans, 1986) | MNL | Downtown, Maastricht, NL | Adults | Outdoor | 426 | 426 | - | - | x | - | - | x |
| (Seneviratne and Morrall, 1985) | Statistics only | Downtown, Alberta, CA | Adults | Outdoor | 2,900 | 2,900 | - | - | x | x | x | - |

**Appendix 2. Overview of studies featuring pedestrian wayfinding performance**

| Author | Field/ Lab/XR | Location of study | Performance variable(s), | Model type | Horizonal / Vertical | Socio-demo | Personal | Complexity | Route | Infra | Network | Info – signs & map | Info –light / sound / color |
|---|---|---|---|---|---|---|---|---|---|---|---|---|---|
| (Feng et al., 2022a) | VR | University building | TD, SP | Statistics | Hor + Ver | - | - | x | - | - | - | - | - |
| (Chen et al., 2023) | VR | Maze | NVF, RS, DGP | Statistics | Hor | - | - | - | - | - | - | x | - |
| (Pouyan et al., 2021) | Field | Hospital | TT, TD, TP, SU, DevTot | Statistics | Hor | x | x | x | - | - | - | - | - |
| (Zhang et al., 2021) | VR | Mall | TP | Trends | Hor | - | - | - | - | x | - | - | - |
| (Shi et al., 2021a) | VR | Maze | TD, NVF, EFF | Statistics, Trends | Hor | - | - | - | - | - | - | x | - |
| (Zhou et al., 2020) | Field | Subway station | EFF | Trends | Hor + Ver | x | - | - | - | - | - | - | - |
| (Kuliga et al., 2019) | Field | Library | TT, TD, NP, TotDev, | MLR | Hor + Ver | - | x | x | - | - | - | - | - |
| (Verghote et al., 2019) | Field | Building | TT, TD, TotDev | Statistics | Hor + Ver | - | - | - | - | - | - | x | - |
| (Li et al., 2019) | VR | Mall | TD, | Statistics, MLR | Hor + Ver | - | - | - | - | x | - | - | - |
| (Cao et al., 2019) | VR | Museum | TT, TD | Statistics | Ver | - | - | - | - | - | - | - | - |
| (Yang et al., 2018) | VR | Maze | ACC | Statistics | Hor | - | - | - | - | - | - | - | - |
| (Slone et al., 2015) | VR | Building | ACC, D | Statistics | Hor | x | - | x | - | - | x | - | - |
| (Vilar et al., 2014a) | VR | Indoor building | TD, TT, NP, SP | Statistics | Hor | x | x | - | - | - | - | x | - |
| (Taillade et al., 2013) | VR | Bordeaux district | ACC, NP | Statistics, trends | Hor | x | x | - | - | - | - | - | - |
| (Lee and Kline, 2011) | VR | Hospital | ACC | Statistics | Hor | x | - | - | - | - | - | x | - |
| (Heino et al., 2010) | VR | Building | TD | Statistics | Hor | - | - | - | - | - | - | x | - |
| (Chen et al., 2009) | VR | Aquatic world | TT | Statistics, trends | Hor + Ver | x | - | - | - | - | - | x | - |
| (Cliburn and Rilea, 2008) | VR | Maze | TD, TT, Y | Statistics, trends | Hor | - | - | - | - | - | - | x | - |
| (Goldiez et al., 2007) | VR | Maze | TT, ACC | Statistics | Hor | - | - | - | - | - | - | x | - |
| (Hölscher et al., 2007) | Field | Conference centre | TT, NP, TD, EFF, SP | Statistics | Hor + Ver | - | x | x | - | - | - | - | - |
| (Morganti et al., 2007) | VR | Building | ACC | Statistics | Hor | - | - | - | - | - | - | - | - |
| (Omer and Goldblatt, 2007) | VR | Urban environment | TD, EFF, TC | Descriptives | Hor | - | - | x | x | - | - | - | - |
| (Soh and Smith-Jackson, 2004) | Field | Forest, Blacksburg, USA | TT, Acc, DM, TotDev | Statistics, Simultaneous regression | Hor | x | x | - | - | - | - | x | - |
| (Jansen-Osmann and Wiedenbauer, 2004) | VR | Maze | NT, NTr, TD, MA | Statistics | Hor | x | x | - | - | - | - | - | - |
| (Kato and Takeuchi, 2003) | Field | Residential area | ACC | Descriptives, Statistics, MLR | Hor | - | x | - | - | - | - | - | - |
| (Malinowski and Gillespie, 2001) | Field | Woodland terrain | TT, TC, ACC | Statistics | Hor | x | x | - | - | - | - | - | - |
| (Soeda et al., 1997) | Field | Department store, university buildings | ACC | Statistics | Hor + Ver | - | x | - | - | - | - | - | - |
| (O'Neill, 1992) | VR | Building | TT, ACC | Statistics | Hor | - | x | - | - | - | - | - | - |
| (O'Neill, 1991) | Field | Campus buildings | SP, BT, ACC, NP | Statistics, Trends | Hor | - | - | - | - | - | x | x | - |
| (Peponis et al., 1990) | Field | Hospital | FU, TotDev | Statistics | Hor | - | - | - | - | x | - | - | - |
| (Moeser, 1988) | Field | Hospital | ACC | Trends | Hor | - | - | - | - | - | - | - | - |
| (Mackay Yarnal and Coulson, 1982) | Field | National park | ? | ? | Hor | - | - | - | - | - | - | x | - |

* where TT = Total Time, D = Delay, TD = Total Distance, ACC = Accuracy / wrong turns / errors, NT = number of turns, NTr = number of trials, NP = number of pauses, SP = Speed, Y = Yaw/degrees of rotation, Eff = TT/min distance. DM = time to make decision, TotDev = total deviation, MA = mean deviation angle, BT = Backtracking, FU = frequency of use of corridor/line/street, TC = Task completion, TP = Turn preference, NVF = Number of victems found, RS = Rescue Score, DGP = distance to goal path.

**Appendix 3. Overview of studies featuring pedestrian observation behavior**

| Author | Field/Lab/XR | Experimental scenario | Horizonal / Vertical | Socio-demo | Personal | Complexity | Route | Infra | Network | Information |
|---|---|---|---|---|---|---|---|---|---|---|
| (Duarte et al., 2014) | VR | Company headquarters | Hor | - | - | - | - | - | - | x |
| (Suzer et al., 2018) | VR | Airport | Hor | - | - | - | - | - | - | x |
| (Cai et al., 2018) | MR | University building | Hor | - | - | - | - | - | - | x |
| (Tian et al., 2019) | VR | Hotel | Hor | - | - | - | - | - | - | x |
| (Wang et al., 2019) | Field | Shopping mall | Hor | x | - | - | - | - | - | - |
| (Zhang and Park, 2021) | VR | Shopping mall | Hor | - | - | - | - | x | - | - |
| (Feng et al., 2022a) | VR | University building | Hor + Ver | - | - | x | - | - | - | - |

**Appendix 4. Overview of impact of independent variables on route choice, wayfinding performance, head rotation, and hesitation (P means positive impact, N means negative impact).**

| Variable | Type | Route choice | Wayfinding performance | Hesitation | Head rotation |
|---|---|---|---|---|---|
| Dist_tot | Route | N | P | | |
| Dist_avg_straight | Route | P | P | N | |
| Dist_longeststretch | Route | P | P | | |
| Level_no | Route | P | P | | |
| Turns_tot | Route | | P | | |
| Turns_left | Route | | | P | |
| Turns_right | Route | | | N | |
| Overlap | Route | N | | | |
| Ratio_wide | Infrastructure | | N | | |
| Floorsigns | Infrastructure | P | | P | P |
| Firedoor | Infrastructure | | | | N |
| Window | Infrastructure | P | N | | |
| Age | Participant | | | | N |
| Education_Sec | Participant | | N | | |
| Education_BSc | Participant | | P | | P |
| Education_MSc | Participant | | | P | |
| Gender_male | Participant | | P | N | |
| Orientation_bad | Participant | | P | | |
| Orientation_good | Participant | | | | P |
| Gaming_often | Participant | | P | N | |
| VR_sometimes | Participant | | | N | |
| Height | Participant | | P | N | |
| Task_2 | Task | | | N | N |
| Task_3 | Task | | | N | N |